\documentclass[12pt,preprint]{aastex}

\newcommand{\eg}{{e.g.,}}

\slugcomment{}

\shorttitle{The MESS catalog}
\shortauthors{Laag et al.}

\begin{document}

\title{The Multi-wavelength Extreme Starburst Sample of Luminous Galaxies Part I: Sample Characteristics}

\author{Edward Laag\altaffilmark{1}, Steve Croft\altaffilmark{2}, Gabriela Canalizo \altaffilmark{3}, Mark Lacy \altaffilmark{4}}


\altaffiltext{1}{Space Science Applications Laboratory, The Aerospace Corporation, \
P.O. Box 92957, Los Angeles, CA 90009}

\altaffiltext{2}{Dept. of Astronomy, University of California, Berkeley, \
 601 Campbell Hall, Berkeley, CA 94720-3411.}

\altaffiltext{3}{Dept. of Physics and Astronomy, University of California, Riverside, \
900 University Ave., Riverside, CA 92521.}

\altaffiltext{4}{National Radio Astronomy Observatory, \
Charlottesville, VA}

\begin{abstract}
This paper introduces the Multi-wavelength Extreme Starburst Sample (MESS), a new catalog of 138 star-forming galaxies (0.1 $<$ $z$ $<$ 0.3) optically selected from the SDSS using emission line strength diagnostics to have high absolute $SFR$ (minimum 11 $M_{\odot}$ $yr^{-1}$ with median SFR $\sim$ 61 $M_{\odot}$ $yr^{-1}$ based on a Kroupa IMF).  The MESS was designed to complement samples of nearby star-forming galaxies such as the luminous infrared galaxies (LIRGs), and ultraviolet luminous galaxies (UVLGs).  Observations using the multiband imaging photometer (MIPS; 24, 70, and 160$\mu m$ channels) on the \textit{Spitzer Space Telescope} indicate the MESS galaxies have IR luminosities similar to those of LIRGs, with an estimated median $L_{TIR}$ $\sim$ $3\times10^{11}$ $L_{\odot}$.  The selection criteria for the MESS suggests they may be less obscured than typical far-IR selected galaxies with similar estimated SFRs.  20 out of 70 of the MESS objects detected in the GALEX $FUV$ band also appear to be UV luminous galaxies.  We estimate the SFRs based directly on luminosities to determine the agreement for these methods in the MESS.  We compare to the emission line strength technique, since effective measurement of dust attenuation plays a central role in these methods.  We apply an image stacking technique to the VLA FIRST survey radio data to retrieve 1.4 $GHz$ luminosity information for 3/4 of the sample covered by FIRST including sources too faint, and at too high a redshift, to be detected in FIRST.  We also discuss the relationship between the MESS and samples selected through alternative criteria.  Morphologies will be the subject of a forthcoming paper.
\end{abstract}


\section{Introduction}

\subsection{Starburst Galaxies} \label{intropart1}

Some of the most fundamental unanswered questions in cosmology concern the nature of star formation in galaxies, and its relationship to galaxy evolution.  Mounting evidence shows we live in an epoch of relative quiescence in terms of star formation.  A measured star formation rate (SFR) of just 1.0 $M_{\odot}$ $yr^{-1}$ for a present day galaxy would be high compared to SFR density estimates for $z$ $=$ 0 \citep[\eg][]{2001ApJ...558L..31H,2004MNRAS.351.1151B}.  The cosmic SFR density is thought to have reached a maximum level between $z$ $\sim$ 2 -- 3, where galaxies along the Hubble sequence formed the bulk of their stars \citep[\eg][]{2003ApJ...587...25D}.  During that same period, merger rates are also thought to have peaked, making it the epoch of most rapid galaxy evolution \citep[\eg][]{2003AJ....126.1183C}.

In spite of their rarity, prodigiously star-forming galaxies called ``starburst galaxies'', with SFRs ranging from $\sim$ 5 to more than 200 $M_{\odot}$ $yr^{-1}$, can be identified locally through a variety of techniques.  Early references to these objects are found in \citet{1979ARA&A..17..477R} and \citet{1981ApJ...248..105W}.  Starburst galaxies have been defined in multiple ways over the years, including on a per unit mass basis known as specific SFR \citep[\eg][]{2009ApJ...706..599L}.  In this paper we limit our focus to objects with high absolute SFR, since these objects are expected to have luminosities characteristic of (U)LIRGs.  Once thought to be unusual, these galaxies provide us with a window on past epochs when they were the dominant hosts of star formation.  The key to finding them is identifying wavelengths dominated by young stellar populations.

The traditional methods used to estimate SFRs are based on direct measurements of luminosity at various wavelengths.  A review of these techniques is found in \citet{1998ARA&A..36..189K}.  However, the focus of that review is on normal galaxies along the Hubble sequence, not the more extreme star forming objects.  The measure of star formation considered to be least affected by dust extinction is the 1.4 $GHz$ luminosity, which primarily traces synchrotron radiation from type II SN's \citep[\eg][]{1992ARA&A..30..575C,1998ApJ...507..155C}.  Using 1.4 $GHz$ luminosities as a reference, many authors have attempted to determine conversion factors for other wavelengths \citep[\eg][]{2003ApJ...599..971H}.  Because of the observed strong far-IR -- radio correlation, the next best estimator is the far-IR luminosity which primarily traces reprocessed UV light from young hot O and B stars.  The conversion from far-IR luminosity to SFR is made more complicated by the presence of an underlying older stellar population \citep{1998ARA&A..36..189K}.  Scatter is generally found to be higher between SFRs based directly on $L_{H\alpha}$, which traces gas ionized by young stars (i.e. HII regions), or $L_{UV}$ produced by young stars directly, and longer wavelengths \citep{1998ApJ...507..155C}.  These conversions are most accurate for the case of vigorously star-forming galaxies with little or no AGN contribution.

\subsection{Far-IR Selection} \label{intropart2}

The {\it Infra-Red Astronomy Satellite} mission \citep[IRAS;][]{1984ApJ...278L...1N} completed the first resolved mid-IR and far-IR survey of the sky, and thereby generated a large catalog of relatively low redshift and dusty star-forming galaxies.  These are known as the Ultra-Luminous Infrared Galaxies (ULIRGs), defined as having $L_{IR}$ $=$ $L(8-1000\mu m)$ $>$ $10^{12}$ $L_{\odot}$, and the factor of $\sim$10 less luminous LIRGs, with $L_{IR}$ $>$ $10^{11}$ $L_{\odot}$\footnote{The cosmology used throughout this paper is $H_{0}$ $=$ 71 $km$ $s^{-1}$ $Mpc^{-1}$, $\Omega_{m}$ $=$ 0.27, and $\Omega_{\Lambda}$ $=$ 0.73.}.  Hereafter we refer to both classes of objects collectively as ``(U)LIRGs'' in most situations, while the terms LIRG and ULIRG will refer to the specific luminosity class defined above.  The {\it Infrared Space Observatory} \citep[ISO;][]{1996A&A...315L..27K} satellite also made significant contributions to their numbers.  At $z$ $<$ 0.1, the number density ($\Phi$) of these galaxies is estimated to be between $10^{-6}$ and $10^{-7}$ $Mpc^{-3}$ for ULIRGs, and between $10^{-4}$ and $10^{-5}$ $Mpc^{-3}$ for LIRGs \citep{1996ARA&A..34..749S}.


Many of the galaxies identified in these surveys continue to be studied, and provide our basic understanding of what (U)LIRGs are, and what physical processes underly their enormous IR luminosities.  Examples include the Revised Bright Galaxy Sample \citep[RBGS;][]{2003AJ....126.1607S} and related Great Observatories All-sky LIRG Survey \citep[GOALS;][]{2009PASP..121..559A}.  A notable sample taken from {\it IRAS} is the ``1 Jy sample'' of 118 ULIRGs, described in \citet{1999ApJ...522..139V} and \citet{2002ApJS..143..315V}.  The 1 Jy sample includes slices of the different forms of activity associated with the (U)LIRG phenomena including Seyfert 2, LINERS, and HII-like galaxies.  (U)LIRGs are also the subject of some excellent review papers \citep[\eg][]{1996ARA&A..34..749S,2006asup.book..285L}.  

(U)LIRGs are found to be predominantly powered by star-formation, but with increasing contributions from AGN for more luminous objects \citep{1999ApJ...522..113V,2007ApJ...656..148A}.  (To distinguish them from AGN dominated ``warm'' sources, starburst powered (U)LIRGs are sometimes referred to as ``cool''.)  They contain significant amounts of dust, and emit as much as 98\% of their total flux in the IR.  For the case of a pure starburst powered (U)LIRG, the predicted dust temperatures range from 30 -- 60 $K$, leading to peak blackbody IR emission of between 60 -- 80 microns \citep{1996ARA&A..34..749S}.  In this sense, the far-IR band has been compared to a calorimeter that gauges star formation activity.

Follow-up observations of sub-mm galaxies (SMGs) at other wavelengths, such as those by \citet{2002MNRAS.331..839F} and \citet{2004MNRAS.355..485B}, have determined these sources are possible high redshift ($z$ $\sim$ 1.5 -- 2.5) counterparts to local ULIRGs, albeit with much higher space densities, possibly several hundred per square degree, and predicted SFRs of $\sim$ 300 $M_{\odot}/yr$ or more \citep{2003MNRAS.344..385B}.  Compared to the evolution of SFR density in the Universe over this period, the change in the density of ULIRGs is dramatic.

\subsection{UV and Optical Selection} \label{intropart3}

While far-IR selection is notable for its success at finding a large population of unknown sources, techniques at shorter wavelengths have also been used to identify objects with intense star forming activity.  The first is the Luminous Compact Blue Galaxies \citep[LCBGs;][]{1997ApJ...489..543P,2004ApJ...615..689G,2007ApJ...671..310G} identified through surveys in the I-band, and selected for their unusually high surface brightness.  As their name suggests, they have a much bluer color than a typical (U)LIRG, with $B-V$ $<$ 0.6 \citep{2008ASPC..396..431P}.  They have median stellar masses of $M_{*}$ $\sim$ $5\times$ $10^{9}$ $M_{\odot}$ and average $E(B-V)$ $\sim$0.5 \citep{2003ApJ...586L..45G}, meaning they have a low mass and are not very extincted.  A related type is the HII or Blue Compact Dwarf galaxies, which are low stellar mass blue starburst galaxies \citep{2005ApJS..156..345G}.  Like (U)LIRGs, LCBGs and HII galaxies are found to be rare locally.  However, by $z$ $\sim$1, LCBGs become ten times more common, and are thought to contribute up to 45\% of the star formation rate density (SFRD) in the Universe \citep{2008ASPC..396..431P}.  The estimated SFR for a typical LCBG may be as high as 40 $M_{\odot}$ $yr^{-1}$ \citep{2001ApJ...550..570H}.

A successful technique used to discover high redshift starbursts is the Lyman break method which relies on the strong attenuation of wavelengths shorter than the Lyman limit (rest frame 912\AA) \citep{1993AJ....105.2017S}.  Follow up observations of these Lyman Break Galaxies (LBGs) determined that they are high-$z$ ultra-violet (UV) luminous star-forming galaxies with moderately high SFRs \citep[\eg][]{1995AJ....110.2519S,1999ApJ...519....1S}, and $M_{*}$ of $10^{9.5}$ -- $10^{11.0}$ $M_{\odot}$ \citep{2002ARA&A..40..579G}.  They have received great attention due to their abundance at high redshifts, and the fact that they may be candidates for progenitors of present day elliptical galaxies  \citep{2002ARA&A..40..579G,2005ApJ...619..697A}.  Some of the most distant galaxies known are the Lyman Alpha Emitters \citep[LAEs;][]{1998AJ....116.2617S,2004ApJ...617..707D}.  These are also thought to be proto-galaxies similar to the early Milky Way; however their true nature is still a matter of debate.

\citet{2005ApJ...619L..35H} have used {\it GALEX} \citep{2005IAUS..216..221M} observations to show there exists a nearby population of UV luminous galaxies (UVLGs) that have strikingly similar properties to LBGs.  \citet{2007ApJS..173..441H} describes a sample of 215 relatively nearby UVLGs that overlap with SDSS.  These have a range of SFRs from a few, to as much as 138 $M_{\odot}$ $yr^{-1}$.   A large portion of these are similar to the LCBG and HII galaxies mentioned above, however a subset of 42 ``supercompact'' UVLGs described by \citet{2007ApJS..173..457B} are thought to be local LBG analogues.  

Finally, the spectroscopic surveys targeting large numbers of galaxies like the 2dF Galaxy Redshift Survey \citep[2dFGRS;][]{2001MNRAS.328.1039C} and the Sloan Digital Sky Survey \citep[SDSS;][]{2000AJ....120.1579Y} allow selection of starburst galaxies via the $H\alpha$ emission line, or from fits to the whole optical spectrum \citep[\eg][]{2007MNRAS.381..494O}.  We discuss a recent survey of this type using SDSS in Section \ref{sample}.

\subsection{Deep Surveys}  \label{intropart4}

Deep surveys targeting objects at high redshift -- combining data sets from {\it HST}, {\it Spitzer}, and {\it Chandra}, with ground based observations -- have brought about a new era of research.  The large projects that have contributed significantly to our knowledge of starburst galaxies include COSMOS \citep{2007ApJS..172...38S}, GOODS \citep{2003mglh.conf..324D}, and CDFS \citep{2004A&A...421..913W} among others.  For example, \citep{2005ApJ...631L..13D} used GOODS-North to select the $BzK$ sample of star-forming galaxies at $z$ $\sim$ 2.  The Multi-band Imaging Photometer \citep[MIPS; 24, 70 and 160$\mu m$ channels;][]{2004ApJS..154...25R} aboard the \textit{Spitzer Space Telescope} \citep{2004ApJS..154....1W}, was a crucial tool for measuring the IR luminosity of distant sources and confirming their status as (U)LIRGs. 

A better picture is developing of where star-formation is occurring in the Universe over time.  As early results from sub-mm surveys alluded to, beyond $z$ $\sim$ 1 star formation is predominantly occurring in (U)LIRGs with major contributions from LBGs and objects similar to LCBGs.  \citet{2006ApJ...637..727C} found that by about $z$ $\sim$1 the mid-IR luminosity function is dominated by LIRGs with stellar masses between $M_{*}$ $\sim$ $10^{10}$ -- $10^{11}$ $M_{\odot}$.  \citet{2005ApJ...631L..13D} find that by $z$ $\sim$2, the typical galaxy with an $M_{*}$ $\sim$ $10^{11.0}$ $M_{\odot}$ is a ULIRG with $L_{IR}$ $>$ $10^{12}$ $L_{\odot}$ and SFR $\sim$ 200 -- 300 $M_{\odot}/yr$.  The co-moving density may be as much as a factor of 1000 greater than the local density.  

Starburst galaxies have been shown to exist in large enough numbers to account for the bulk of star-forming activity in the early Universe.  As the large surveys push detailed observations to higher and higher redshifts, it is becoming increasingly important to understand the complex relationships between various star formation rate indicators.  Clearly it is not sufficient to rely only upon samples of ``typical'' or ``well behaved'' galaxies observed today.  Though they present unique observational challenges, one of which is the frequent high levels of dust obscuration, it is worthwhile to probe these difficult objects.  

\subsection{Previous Work} \label{intropart5}

Over the years many attempts have been made to derive SFRs from the $H\alpha$ emission line luminosity directly, and to explore the relationship between UV, far-IR, and radio derived SFRs.  The SFRs based on uncorrected $L_{H\alpha}$ are generally lower than SFRs measured from the IR.  \citet{2002AJ....124.3135K} (their figure 1) find a correlation between uncorrected $SFR_{H\alpha}$ and $SFR_{IR}$ (computed using formulae in \citet{1998ARA&A..36..189K}), but that the $SFR_{H\alpha}$ underestimates by about a factor 3 the $SFR_{IR}$.  They find the amount the $H\alpha$ underestimates the SFR increases for galaxies with higher SFRs.  After correcting $L_{H\alpha}$ using a Balmer decrement derived E(B-V), they find much better agreement between the two methods .  However, their sample is drawn from the Nearby Field Galaxy Survey, which is composed of less vigorously star-forming galaxies than typical (U)LIRGs.  It includes only 1 object with IR or corrected $SFR_{H\alpha}$ greater than 50 $M_{\odot}$ $yr^{-1}$, and only three with IR or corrected $SFR_{H\alpha}$ greater than 20 $M_{\odot}$ $yr^{-1}$.

\citet{2006ApJ...637..227C}, using data from the {\it Spitzer} First Look Survey, compute an optical SFR derived from emission lines for a sample which includes a significant number of LIRGs with IR predicted SFRs between about 20 and 105 $M_{\odot}$ $yr^{-1}$.  None of their uncorrected optical SFRs are more than 20 $M_{\odot}$ $yr^{-1}$.  They find scatter at IR luminosities greater than $10^{10}$ $L_{\odot}$ (their figure 9).  Finally, \citet{2004A&A...415..885F} examine a sample of $ISO$ selected LIRGs, nine of which have $H\alpha$ derived SFR greater than 20 $M_{\odot}$ $yr^{-1}$ (their figure 2b). They find a non-linear relationship between the corrected $SFR_{H\alpha}$ and $SFR_{IR}$ (to which they fit a polynomial) which increases as IR luminosity increases.  


It is strongly suggested there should be a physical connection between optical extinction, higher SFR, dustier galaxies, and higher $L_{TIR}$.  \citet{1996ApJ...457..645W} observed that the UV/FIR ratio decreases with increasing FIR luminosity.  Other authors have found a correlation, albeit weak, between SFR and extinction measured by the $H\alpha/H\beta$ ratio, inferring dustier galaxies will generally have higher SFR \citep[][figure 4]{2001ApJ...558...72S}.

\clearpage

\section{Sample Selection} \label{sample}

We have taken a different approach to selecting our sample of starburst galaxies, with the aim of finding nearby objects having SFRs at or above the (U)LIRG level, but with less dust obscuration.  In addition to lending themselves to more detailed study at shorter wavelengths, objects in this sample have the potential of being intermediate (or transitioning) objects between categories mentioned above.  They can also be used to explore the relationships between various SFR indicators.  Rather than relying on a single emission line, the goal of our selection method is to identify galaxies whose entire spectrum indicates an unusually high level of star-forming activity.  For this reason we decided to use the already available SDSS catalog of SFRs from \citet{2004MNRAS.351.1151B}, described below, as a starting point.

\citet{2004MNRAS.351.1151B} presented measured SFRs for a sample of $\sim$53,000 star-forming galaxies observed by the SDSS (henceforth we refer to \citet{2004MNRAS.351.1151B} as ``B04'').  B04 employ a novel technique to determine SFRs, rather than the fixed conversion factor estimators like those in \citet{1998ARA&A..36..189K}.  In short, they apply galaxy evolution and emission line modeling to generate model grids corresponding to galaxy-wide parameters, and given the emission line spectrum, compute a likelihood that a given model is correct.  The value of the most likely SFR for each source optical spectrum is then independent of UV, IR, and radio properties.

Rather than using the default SDSS spectroscopic pipelines, B04 use their own optimized pipeline to re-analyze the 1D spectra.  This data set is now known as the MPA/JHU value-added galaxy catalog.  The data reduction for this catalog is described in more detail by \citet{2004ApJ...613..898T} .   The benefit of using this data set over the standard SDSS pipeline, is the improved accuracy in continuum subtraction.  This results in much better identification of emission lines, particularly the weaker (low SNR) ones.  This precision is critical for performing the various tests to identify and remove AGNs described in section \ref{bpt}.

B04 build upon methodology outlined in \citet{2002MNRAS.330..876C}, modeling the emission lines following \citet{2001MNRAS.323..887C}, and with galaxy evolution models from \citet{1993ApJ...405..538B}.  Their model grids take into account parameters such as metallicity, ionization parameter, and dust attenuation.  These grids contain $\sim$ $2\times10^{5}$ models.  Each model in the grid has an associated dust attenuation based on all the emission lines, however B04 state: ``To first approximation, however, our dust corrections are based on the $H\alpha$/$H\beta$ ratio.''  Then comparing to the data, they use a Bayesian approach to compute a likelihood for each model.   In this manner, a likelihood distribution for the value of SFR is generated spanning a moderate range of SFRs.

As of January 2010, the MPA/JHU catalogs, including the SFRs were updated to include a large number of additional galaxies from SDSS Data Release 7 (DR7).    At this time, they also implemented several refinements to their data reduction pipeline and SFR estimation techniques.  Significantly, the stellar masses are now calculated using fits to photometry and the aperture correction method has been changed to remove a systematic overestimate of total SFR for certain galaxy classes as identified by \citet{2007ApJS..173..267S}  -- (see the web page: \textit{http://www.mpa-garching.mpg.de/SDSS/DR7/index.html} for details).  Our original sample selection (described below) relied on the Data Release 4 (DR4) version of the MPA/JHU catalogs, however, throughout this paper we use the updated DR7 values for all of our analysis.

In querying the B04 SFR catalogs, we chose to select the average value of the likelihood distributions.  Since the SDSS fibers are relatively small, an aperture correction based on color information is applied to the values.  Two distributions are thus generated by B04, one corresponding to the fiber magnitude, and the other to the total magnitude.  These average values of the distribution become what we will call the ``\emph{fiber} SFR'' and the ``\emph{total} SFR''.  The same applies for the stellar mass, which we also extract from the catalog.  Additionally, we extract the 16 and 84 percentiles of the likelihood distributions as a measure of uncertainty.  Finally, we extract from the catalogs the gas phase oxygen abundances determined by \citet{2004ApJ...613..898T} in units of 12$+$ $log$(O/H), as a measure of the metallicity.

In selecting the MESS, we used these B04 SFR estimates to aid in identifying objects with the potential to be starburst galaxies; however, we do not rely solely upon these as the definitive SFR.  In Section \ref{sfr} we compare the B04 SFR to other more traditional methods.  It is also important to mention that B04 use a Kroupa IMF.  It is possible to convert their SFR to a Salpeter equivalent by multiplying the B04 SFR by a factor of 1.5.

To generate the MESS, we queried the original B04 database (based on SDSS Data Release 4 at the time) with the following criteria:

\begin{enumerate}
 \item SFR $>$ 50$M_{\odot}/yr$ for both \emph{total} SFR and \emph{in fiber} values
 \item No excessive corrections from \emph{fiber} to \emph{total} SFR or stellar mass (see below)
 \item SNR $>$ 3 detection on all emission lines (class $=$ 1 objects in B04)
 \item 0.1 $<$ $z$ $<$ 0.3 
\end{enumerate}

We imposed criteria 1 and 2 to help ensure our sample was not dominated by spuriously large corrections on the fiber values.  A small number of objects in the B04 sample had absurdly large total SFRs, caused by extreme aperture corrections. Often these occurred when the color corrections failed because the fiber was centered directly on a small galaxy, but there was a nearby bright star contributing a large amount of blue light. We excluded such cases by rejecting all objects with aperture corrections larger than a factor 30. The aperture corrections for objects which survived this cut were generally small, and as of DR7 did not exceed a factor of 2 from fiber to total SFR.  The majority of the corrections on the masses from fiber to total were a factor of 3 or less, with only a handful of objects exceeding this (largest was $\sim$7).  In other words, there are no objects in our sample with corrections anywhere near as large as our conservative cut of a factor 30.  The query criteria, combined with the small 3\arcsec\ spectroscopic fiber size of SDSS, also means the selected objects have high SFR in a relatively compact area.  Criterion 3 ensures that we are able to classify objects as star-forming galaxies or AGN.  Consequently, the MESS contains only emission line galaxies.  Finally, criterion 4 means the sample is relatively low redshift, but still probes a range where there is the potential of discovering many new (U)LIRGs.  It also assures useful emission lines like [SII] will not be redshifted out of the optical spectra.  The 138 objects that met these search query criteria form a complete sample within the SDSS DR4 footprint; they are listed in Table \ref{tab:tableone}.

The 2010 update to the 2004 catalogues resulted in significant revisions to the MESS SFRs.   In some cases (40) the values were revised below the original threshold of 50.  However, the median total SFR (Kroupa IMF) for the sample remains high at 61 $M_{\odot}/yr$, and all but three of the revised optical SFRs are still high enough that they would satisfy the minimum LIRG level SFR indicated by the IR Kennicutt relation (20 $M_{\odot}/yr$).  See Appendix A for further details, and a histogram demonstrating changes from DR4 to DR7.  The total SFRs now  range from a minimum of 11 $M_{\odot}/yr$, to one object which has an estimated 200 $M_{\odot}/yr$.   The median estimated $log$ ($M_{*}$/$M_{\odot}$) (stellar mass) value from the catalogue is 11.1, making them moderately massive galaxies.  The median redshift of MESS sources is $0.200$, with a fairly even distribution of $z$ values.

It is worth noting that by selecting a sample with high absolute SFR (at the (U)LIRG level), we have obviously excluded from our sample starburst galaxies with lower SFR and stellar mass, but high SFR relative to their mass (specific SFR).  For example, the sample in \citet{2009ApJ...706..599L} includes a large portion of dwarf galaxies that would be missed by our selection criteria.  Our main goal was to find objects comparable in total luminosity to (U)LIRGs, in terms of optically measured SFR.  However, the objects in our sample do seem to have high specific SFR, as indicated by the B04 stellar masses. See Section \ref{sfr}.

In the remaining sections of the paper we will use the MESS to explore the relationship between various SFR indicators, and between the MESS and samples selected using other methods.  

\section{Multi-wavelength Data}
\subsection{SDSS Photometry} \label{sdssphot}

\begin{figure}
\epsscale{.75}
 \plotone{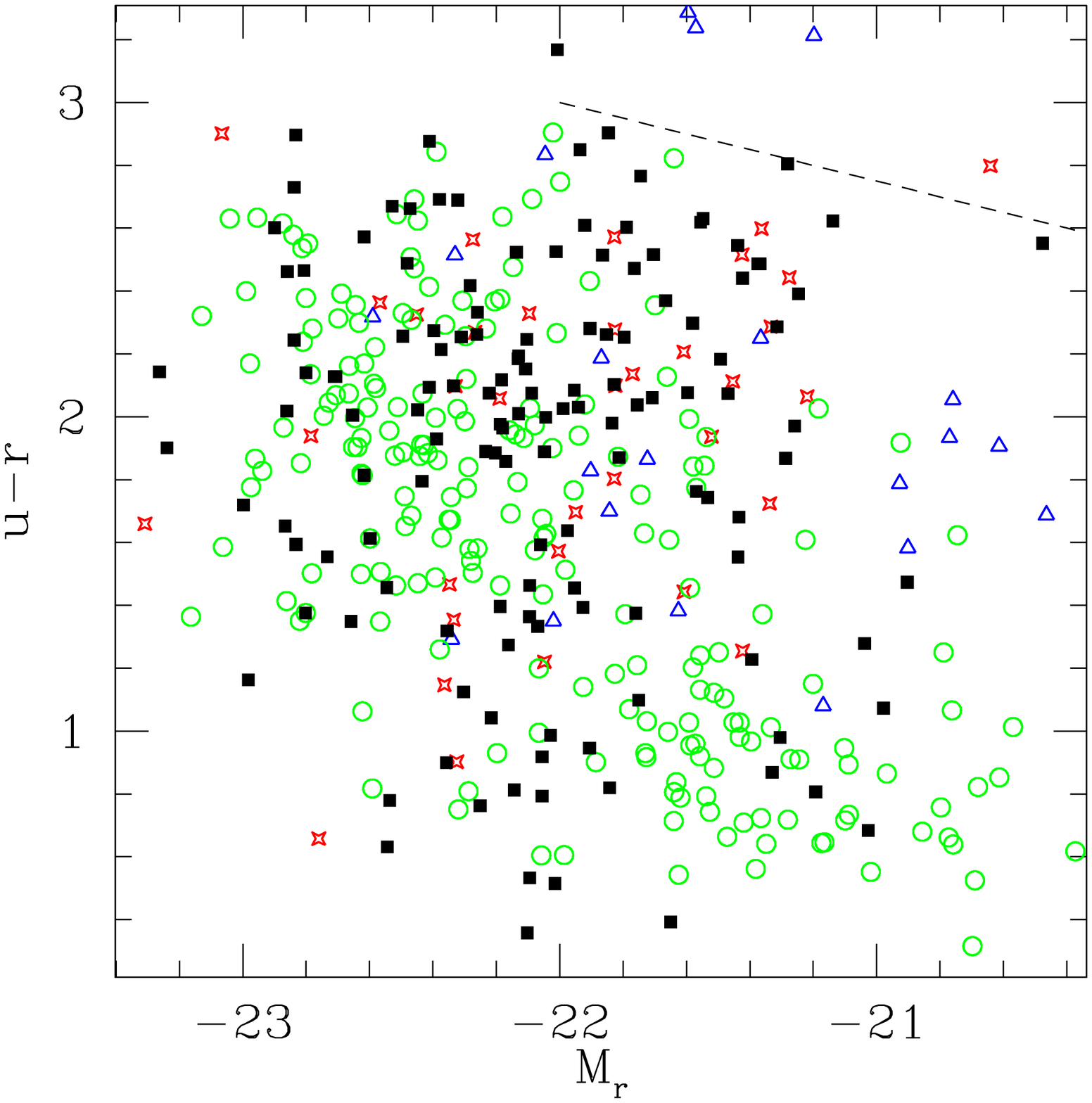}
 \caption{A color-magnitude diagram using photometry from SDSS DR7 for the MESS galaxies.  The filled black squares represent the MESS catalog.  Also plotted are the UVLGs (green circles) sample from \citet{2007ApJS..173..441H}, the 1 Jy sample of ULIRGs (red stars) from \citet{1999ApJ...522..113V} and the FIRST sample of (U)LIRGS (blue triangles) \citet{2000ApJS..131..185S}.  For the $u$ $-$ $r$ color we use model magnitudes, and for the $M_{r}$ we use Petrosian magnitudes.  The dashed line in upper right corner represents the \emph{approximate} location of the ``red sequence'' galaxies at $z$ $=$ 0.  
\label{fig1}}
 \end{figure}

Since the MESS is selected from SDSS \citep{2000AJ....120.1579Y}, we have access to high quality visible imaging/photometry data and spectra.  Figure \ref{fig1} is a color-magnitude diagram of the MESS using SDSS photometry, comparable to similar diagrams in \citet{2004ApJ...608..752B} and \citet{2001AJ....122.1861S}.  Also plotted are galaxies from the UVLGs \citep{2007ApJS..173..441H} sample, and portions of the 1 Jy \citep{1999ApJ...522..113V} and FIRST samples \citep{2000ApJS..131..185S} covered by SDSS.  The dashed line in the upper right region represents the approximate location of the ``red sequence'' at $z$ $\sim$ 0.1, along which early type galaxies tend to cluster.  This figure demonstrates the large range of colors spanned by the MESS, including the ``blue cloud'' through the ``green valley''.

\subsection{Power Source Identification} \label{bpt}

\begin{figure}
\epsscale{.75}
 \plotone{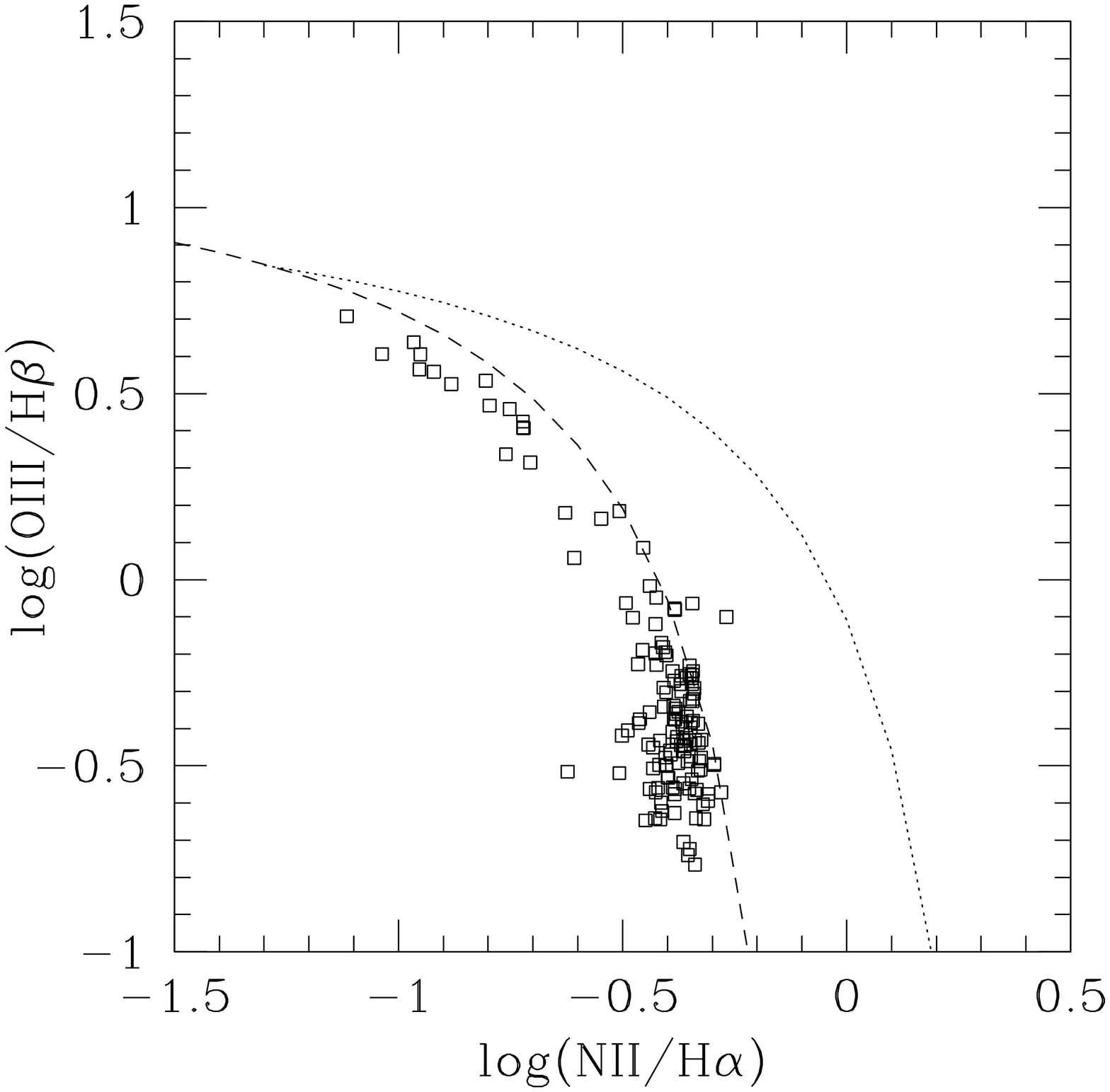}
 \caption{BPT diagram 1 for the MESS.  One of three emission line galaxy classification diagrams developed by \citet{1981PASP...93....5B}.  These have been improved upon by \citet{2001ApJ...556..121K} and \citet{2003MNRAS.346.1055K}.  The dotted line represents a maximal starburst level defined in \citet{2001ApJ...556..121K} and the dashed line the limit for pure star formation defined by \citet{2003MNRAS.346.1055K}.  The area above these lines represents objects mainly powered by some form of AGN.  The region in between these curves is generally thought to represent composite objects.  The region below the curves is occupied by star-forming (HII-like) galaxies.  More information on these and the next two diagrams can be found in \citet{2006MNRAS.372..961K}. The emission line fluxes are taken from the MPA/JHU value-added catalog. \label{fig2}}
 \end{figure}

\begin{figure}
\epsscale{.75}
 \plotone{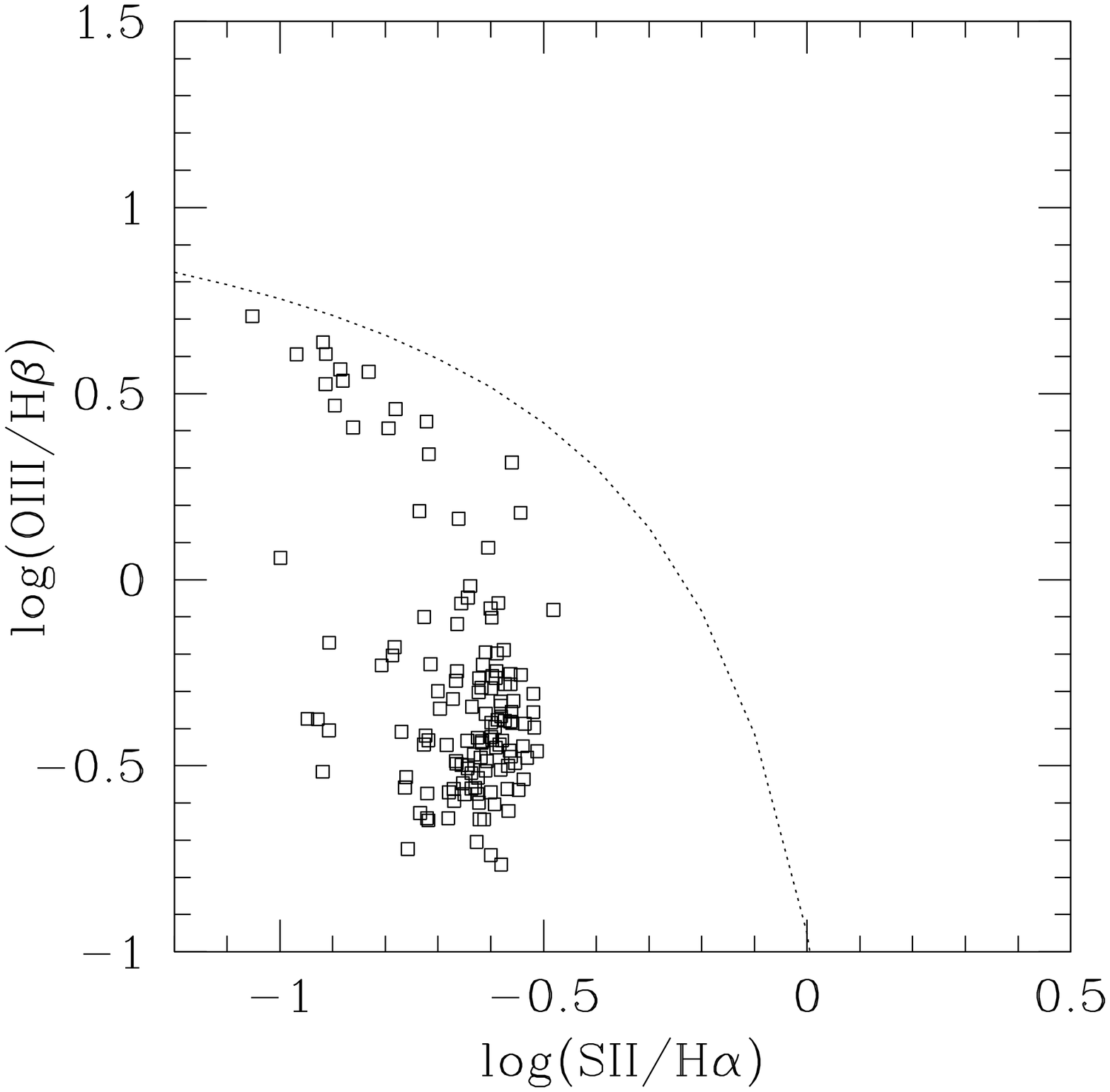}
 \caption{BPT diagram 2 for the MESS.  The dotted line marks the division between star formation and AGN powered objects. \label{fig3}}
 \end{figure}

\begin{figure}
\epsscale{.75}
 \plotone{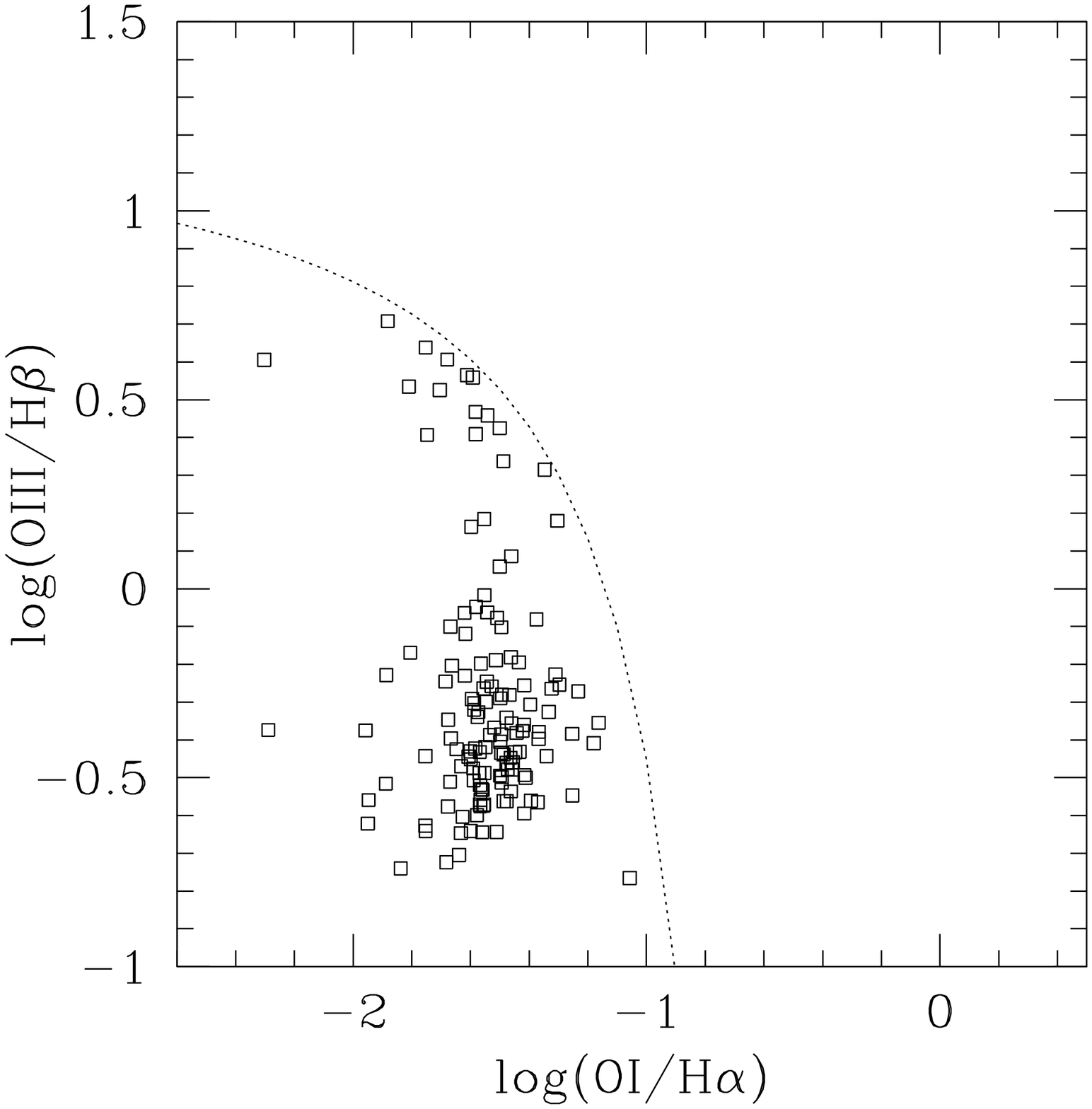}
 \caption{BPT diagram 3 for the MESS. The dotted line marks the division between star formation and AGN powered objects.    
 \label{fig4}}
 \end{figure}

We have used the so-called ``BPT diagrams'' \citep{1981PASP...93....5B}, updated with the improved classification schemes presented by \citet{2001ApJ...556..121K} and \citet{2003MNRAS.346.1055K}, to verify that our galaxies are powered by star formation in all 3 forms of the diagram.  This ensures the observed luminosities are due primarily to starbursts rather than AGN; however this does not guarantee our objects do not contain a ``buried'' AGN.  (We obtained the emission line fluxes from the DR7 MPA/JHU catalog).  It can be seen from Fig.~\ref{fig2}, Fig.~\ref{fig3}, and Fig.~\ref{fig4}  that our galaxies lie almost wholly beneath the line of pure star formation.  The region occupied by the MESS are also denoted by many authors as ``HII'' type galaxy spectra \citep[\eg][]{2006MNRAS.372..961K}.   We will discuss further the possibility of AGN contamination in sections \ref{farir}, and \ref{radio}, however we believe the implication of the above tests is that AGN are making a minimal contribution, if any, to the IR luminosity of the MESS objects.

Additional information can be gleaned from the [OIII] to $H\beta$ ratio.  Some authors suggest this ratio is sensitive to recent starburst activity in HII galaxies \citep{2007ApJ...654..226R,2007ApJS..173..457B}.  Furthermore, the equivalent width (EQW) of the $H\beta$ emission line is thought to be a measure of the ratio of present to past star formation, so that recent single starbursts would have both higher [OIII]/$H\beta$ ratio and larger EQW($H\beta$).  We have examined the EQW($H\beta$) and [OIII]/$H\beta$ ratio for the MESS.  While this line is typically weak, $\lesssim$ 11 \AA\ for the MESS, we find 9 objects with $H\beta$ EQWs $>$ 50 {\AA\ }.  These correspond to MESS sources: J004236$+$160202, J020038$-$005954, J074936$+$333716, J095618$+$430727, J115630$+$500822, J145435$+$452856, J150627$+$562702, J154049$+$390350, and J154120$+$453619.  These sources also have higher [OIII]/$H\beta$ ratio, and have blue colors based on the optical photometry.  These objects are more representative of the ``supercompact UVLGs'' identified by \citet{2007ApJS..173..457B} (see their figure 13), than they are of the rest of our MESS catalog.

\subsection{Far-IR Observations} \label{farir}

We obtained space-based observations in order to study the far-IR properties of the MESS, and to compare them to (U)LIRGs in classically-selected samples.  We have acquired data with the Multi-band Imaging Photometer \citep[MIPS; 24, 70 and 160$\mu m$ channels,][]{2004ApJS..154...25R} aboard the \textit{Spitzer Space Telescope} for all 138 MESS objects (Program ID 40640).  The data were obtained in MIPS photometry mode with the exception of 3 sources in scan mode identified below.  These data were automatically processed through the Spitzer Science Center (SSC) data pipelines, with version numbers ranging from 16.1.0 for the earliest data, and up to 18.5.0 for the most recent.

We began our MIPS data reduction with the basic calibrated data (BCD) products.  For the 24 $\mu m$ channel, we flat-fielded the BCDs using the ``flatfield.pl'' script from the SSC.  We then corrected for overlap, and re-mosaicked the BCDs using the MOPEX software package \citep{2005ASPC..347...81M} available from the SSC.  For the 70 and 160 $\mu m$ channels the delivered filtered BCD products showed filtering artifacts due to the presence of bright point sources, particularly for the 160 $\mu m$ channel.  To mitigate this we used scripts delivered with the SSC Germanium Reprocessing Tool (GeRT) software package to filter the regular BCDs while masking out the location of bright point sources.  We then mosaicked these masked and filtered BCDs with MOPEX.  The MOPEX package includes an APEX point source extraction utility \citep{2004AAS...20515312M} that was used to measure aperture photometry for all 3 channels.  We then applied the standard aperture corrections for point source photometry available from the SSC website MIPS data handbook to the measured fluxes (specifically the corrections were for a 7.0\arcsec\ radius aperture at 24 $\mu m$, 16.0\arcsec\ radius and 60$K$ source at  70 $\mu m$, and 16.0\arcsec\ radius with 50$K$ source at 160 $\mu m$).  We report photometry results in Table \ref{tab:tabletwo}.   A color correction has not been applied to these values.  The manner in which we compute infrared luminosity described below assumes an SED incorporating a range of source temperatures, so we have reported the actual values we use for that relation.  It should be noted that the uncertainties reported contain terms added in quadrature to account for stability and calibration errors according to the MIPS Instrument Handbook, in addition to the photometric uncertainty.  These measurement uncertainties become the basis for the typical/average error bar representation on the figures.   Some additional noise terms unique to the instrument such as ``confusion error" have not been included in the uncertainties, but nonetheless, it should be understood these will affect the sample on a case by case basis.

When the angular sizes and distances of the MESS sources are taken into account, and combined with the pixel scales for the MIPS mosaicked images (2\farcs45, 4\farcs00, and 8\farcs00 per pixel respectively), it is not surprising most of our objects appear as point sources in all 3 channels.  In a handful of merger cases, the galaxies were resolved into two sources at 24 $\mu m$.  For those objects the point source aperture fluxes were summed.  At 70 and 160 $\mu m$ these sources are no longer well resolved into two distinct objects.  The majority of the MESS objects do not lie in regions of extended emission or high IR background levels.  At the MIPS wavelengths the sources generally appear quite isolated.

The majority of the MESS galaxies were detected with high SNR by MIPS using the APEX tool.  However, three sources (J022229$+$002900, J040210$-$054630, J150627$+$562702) at 70 $\mu m$, and fifteen sources (J004236$+$160202, J020038$-$005954, J020215$+$131749, J021601$-$010312, J022229$+$002900, J033918$-$011424, J040210$-$054630, J085906$+$542150, J095618$+$430727, J124907$+$582729, J131101$-$004215, J145435$+$452856, J150627$+$562702, J151320$-$002551, J154120$+$453619) at 160 $\mu m$, are either not detected with APEX, or had a SNR for the photometry indicated as $\lesssim$ 5.  For sources that were not detected with APEX, we used IRAF phot to measure the flux density centered at the 24 $\mu m$ source position for the 70 and 160 $\mu m$ measurements (within the appropriately sized aperture).  Although the resulting measurements have large uncertainties, using measured aperture photometry at the position of the object (as determined from the 24um data) provides a truer measure of the source fluxes than simply using a 5$\sigma$ upper limit value for all sources (which will tend to bias their fluxes high).  Finally, there were 3 sources (J104116$+$565345, J104729$+$572842, J235237$-$102943) which were detected in the MIPS scan mode for a previous proposal, and were not re-observed for our program.  

For the majority of the MESS, we only have far-IR data for the three MIPS channels.  However, a subset of 36 sources were also detected by {\it IRAS}.  We discuss these data in section \ref{iras} below.  The other MESS objects are simply below the sensitivity limits for {\it IRAS} (in survey mode).

Using the MIPS data we have calculated the bolometric infrared luminosity, $L_{TIR}$, for each galaxy in the sample.  A traditional method would be to do a simple single temperature modified blackbody fit to the points.  These models have been used to approximate the far-IR SED for a galaxy, but are not physically realistic since the actual IR SED for a star-forming galaxy is built up from a combination of blackbody emission profiles spanning a range of temperatures.  For purposes of computing $L_{TIR}$ we do not need to constrain the exact SED in order to generate reliable estimates; rather, we can simulate the full range of normal star-forming galaxy IR SEDs.  This approach is described in \citet{2002ApJ...576..159D}.  They derive a relation (their equation 4) designed to recover the total infrared (TIR) luminosity for star-forming galaxy SED shapes.  We reproduce the relation here.

\begin{equation}
L_{TIR}=\zeta_{1}*\nu*L_{\nu}(24 \mu m)+\zeta_{2}*\nu*L_{\nu}(70 \mu m)+ \zeta_{3}*\nu*L_{\nu}(160 \mu m)
\end{equation}

Given a range of model SED shapes at $z$ $=$ 0, the formula has been shown to match the bolometric infrared luminosity to a high degree of accuracy (within 4\% to redshift 4).  For our sample and redshift range, we are probably subject to $\sim$ 10\% error (Daniel A. Dale via Priv. Comm.).  We use this method to compute the $L_{TIR}$ for the MESS, applying the appropriate coefficients ($\zeta_{1}$, $\zeta_{2}$, $\zeta_{3}$) for their equation given that our redshifts range from $z$ $=$ 0.1 to 0.3.  We obtained the coefficients from the authors (via priv. comm.).  This method for computing $L_{TIR}$ is similar to a relation derived for $L_{FIR}$ by \citet{1996ARA&A..34..749S} using the $IRAS$ bands.  The latter was commonly used to estimate $L_{FIR}$ for the $IRAS$ selected samples of (U)LIRGs.

Note that the total luminosity measured in the TIR range of 3 - 1100 $\mu m$ is not appreciably different to that measured in the IR range of 8 - 1000 $\mu m$, or even in the smaller FIR range 40 - 500 $\mu m$, since all of these ranges cover the FIR region, where the bulk of emission for a dusty star-forming galaxy will occur.  In the subsequent analysis, and in sections that follow, we will treat the $L_{TIR}$ and $L_{IR}$ as basically indistinguishable at this level, using the TIR subscript to denote only the method used to determine it.  To summarize the properties of the MIPS data for our sample, the median flux at 70 $\mu m$ is 155 $mJy$.  The median $log$ ($L_{TIR}$/$L_{\odot}$) obtained is 11.5.   


In Fig.~\ref{fig5} we plot the resulting $L_{TIR}$ for our complete sample, versus the B04 total SFR (Kroupa IMF).  As explained in section \ref{intropart1}, SFRs are frequently estimated directly from $L_{TIR}$ (see section \ref{sfr}).   There is obviously significant scatter when we compare the two quantities; we discuss the possible causes in section \ref{disc} below.  Asterisks denote the low SNR detections.

Finally, we can create a far-IR color-color diagram using the MIPS bands (Fig.~\ref{fig6}).  Others have used figures such as this to identify potential ``warm'' (U)LIRGs (originally \citet{1994ApJ...436..102L} and see also \citet{2001ApJ...555..719C}).  The MESS is plotted as black squares.  Also plotted are a subset of the GOALS objects (from GOALS data release 1) \citep{2003AJ....126.1607S} for which MIPS fluxes have been released.  For the purposes of this figure, the same color correction applied to the GOALS data has been applied to our fluxes.  The same basic range is seen in both samples.  Some of the MESS exhibit high $\alpha$(70,24) values.  If we set the threshold for warm objects at a level of $-$2.10 and above, then we find that the sources correspond to objects with low SNR detections at 160 $\mu m$ and/or the optically blue objects identified from SDSS colors.  Additionally, the sources identified previously as having high [OIII]/$H_{\beta}$ and large EQW($H_{\beta}$) correspond to higher positions on this diagram.  It appears the portion of the MESS occupying the higher positions may be representative of young dust poor starbursts.  An alternative explanation would be that they are indeed ``warm'' LIRGs containing a buried AGN, but we find no other evidence for this (see section \ref{bpt}). 

\subsection{Comparison to IRAS} \label{iras}

We have 36 coordinate matches in the {\it IRAS} \citep{1984ApJ...278L...1N} catalogs.  Five of them are detected in the point source catalogue (PSC).  The rest are from the faint source catalog (FSC; SNR $>$ 5), or the faint source reject file (FSR; SNR $>$ 3).  The MESS objects are near the detection limit of {\it IRAS}.  The data include high quality (Fqual $=$3) or moderate quality (Fqual $=$2) data for the 60 $\mu m$ channel, but nearly all the 100 $\mu m$ measurements are upper limits only.   However, the {\it IRAS} data provides a useful independent check on the MIPS fluxes.

Figure \ref{fig7} compares the MIPS 70 $\mu m$ fluxes to the IRAS 60 $\mu m$ values for each {\it IRAS} detected MESS source, with the dotted line representing a one-to-one correspondence.  Taking into account the wavelength difference, we find good agreement between these two bands which reside near the peak emission for starbursts.  Figure \ref{fig8} compares the $L_{FIR}$ computed with formulas in \citet{1996ARA&A..34..749S} to the $L_{TIR}$ computed with the MIPS fluxes using the method described above.  Note that the {\it IRAS} $L_{FIR}$ represented here is based on upper limits at 100 $\mu m$.

\subsection{Radio} \label{radio}

48 of the MESS sources are detected at 1.4 $GHz$ in the VLA FIRST survey \citep[Faint Images of the Radio Sky at Twenty Centimeters;][]{1995ApJ...450..559B}.  117 of the total 138 sample RA and DECs fall within the FIRST coverage area.  The detected MESS objects have a median integrated 1.4 $GHz$ flux of only 1.7 $mJy$, making them among the faintest sources detected by the survey.  FIRST has a 1 $mJy$ source detection threshold.  We obtained the integrated fluxes for these sources from the FIRST website\footnote{http://sundog.stsci.edu/} catalog search.  These values were k-corrected assuming $S_{\nu}$ $\varpropto$ $\nu^{-0.8}$, and converted to luminosity in $W/Hz$.

We have examined the FIRST ``cutout'' images of the MESS sources to look for unusual features such as double lobed radio sources that might be indicative of an FRII galaxy.  In all cases the MESS appear to be point sources, with essentially no structure.  Considering the detection limit of FIRST corresponds to $10^{22.5}$ -- $10^{23.5}$ $W$/$Hz$ at the MESS redshifts, anything detected is either a powerful starburst or AGN. 

A well known correlation exists between radio and far-IR luminosity for many star-forming galaxy types \citep[\eg][]{1985ApJ...298L...7H}.  It is believed the correlation may be used to calibrate a SFR for the IR luminosity.  The advantage being that the 1.4 $GHz$ flux is virtually unaffected by dust attenuation and may provide a less biased value for the most heavily obscured galaxies \citep[\eg][]{2003ApJ...599..971H,2001ApJ...558...72S}.  

In order to test the radio-IR correlation for as much of the sample as possible, we have performed image stacking using all 117 FIRST image cutouts (including the detected and non-detected fields) with the IRAF imcombine task.  This technique, described in \citet{2007ApJ...654...99W}, allows luminosity information to be recovered for objects that fall well below the rms noise level.  The technique has been applied previously to samples of quasars \citep{2007ApJ...654...99W} and AGNs \citep{2007AJ....134..457D}.  A median stack of all 117 MESS image cutouts (where the cutouts have been converted to luminosity units) results in a luminosity 9 $\times$ $10^{22}$ $W/Hz$, after correcting for snapshot bias (see \citet{2007ApJ...654...99W}).  We determined the SNR in the stacked image of all cutouts was therefore high enough to allow us to divide the sample into 5 subsample bins.  The objects were sorted in order of increasing IR luminosity, prior to dividing into the bins.  The cutouts for each bin were then median stacked, and the resultant radio luminosities were measured.

We have used the 1.4 $GHz$ luminosities for each of the detected objects, as well as the median stacked data, to compute 1.4 $GHz$ SFRs.  We discuss these results for the MESS in section \ref{sfr}. 

\subsection{Extinction} \label{dust}

Because our sample is optically selected, it is possible the objects are less dust obscured than other objects with similar SFRs, like typical IR selected (U)LIRGs.  As a measure of the dust extinction in this sample we start with the ratio of $H\alpha$ to $H\beta$ emission lines (Balmer decrement), and apply methods from \citet{1994ApJ...429..582C} to calculate the Balmer optical depth, $\tau_{B}$, and then estimate an $E(B-V)$ from this.  We assume a theoretical unreddened $H\alpha$/$H\beta$ ratio of $2.88$ \citep[\eg][]{1989agna.book.....O}.  

Specifically, we apply \citet{1994ApJ...429..582C} equations 2 and 3, with the Balmer optical depth given by:

\begin{equation} 
\tau^{l}_{B}=\tau_{\beta}-\tau_{\alpha}=ln(\frac{H\alpha/H\beta}{2.88})
\end{equation}

and the resulting relationship to the intrinsic $E(B-V)$ for their sample of starburst and blue compact (HII) galaxies was found to be:

\begin{equation}
E(B-V)_{i}=\frac{1.086}{k(H\beta)-k(H\alpha)}*\tau^{l}_{B}\simeq0.935*\tau^{l}_{B}
\end{equation}

where $k(H\beta)-k(H\alpha)$ was obtained from \citet{1979MNRAS.187P..73S}.  Note that the above assumes a simple foreground screen of obscuring dust.  The Balmer optical depth is also thought to be an upper limit on attenuation \citep{2000ApJ...539..718C}.

Using the methods above, the median $E(B-V)$ for the MESS is $0.648$, and the maximum value obtained is $1.04$.  These values are generally lower than those obtained by \citet{2002ApJS..143..315V} for the $IRAS$ 1 Jy sample of ULIRGs, who found the median $E(B-V)$ for HII galaxies, LINERS, and Seyfert 2 galaxies to be 0.80, 1.11, and 1.21 respectively.  The values obtained for the {\it IRAS} Bright Galaxy Sample by \citet{1995ApJS...98..171V} were 1.05, 1.24, and 1.07 for HII galaxies, LINERS, and Seyfert 2 galaxies respectively.  Doing the above analysis for the Hoopes sample of UVLGs, using data from the MPA/JHU catalog, gives a median value of $0.316$.  This indicates a much lower extinction for that sample.

\subsection{GALEX Detections}

A significant number of the MESS objects (70) are detected by the {\it GALEX} mission with SNR greater than 3 in the FUV band ($\lambda_{fuv}$ $\sim$ 1530 \AA).  We used the Multi-mission Archive at Space Telescope (MAST) explorer tool to query the GR4/GR5 data release catalogs.  The majority of these observations are taken from the all sky survey (AIS; $\sim$ 100 $s$ exposure times) or the medium imaging survey (MIS; $\sim$ 1.5 $ks$ exposure times) \citep{2007ApJS..173..682M}.  An additional 7 targets were contained in the deep imaging survey (DIS; $\sim$ 30 $ks$).  The short exposure times of the AIS and MIS mean that only the most luminous objects are detected. 

Using the obtained fluxes we computed an \emph{observed} $L_{fuv}$, incorporating a correction for foreground reddening using the $E(B-V)$ column in the {\it GALEX} catalog (where $A_{UV}$/$E(B-V)$ assumed was 8.24), and we find 20 of our 138 objects meet the minimum criteria of being a UVLG according to the definition used by \citet{2007ApJS..173..441H}, which is $L_{fuv}$ $>$ $2\times10^{10}$ $L_{\odot}$.  Only three of the MESS were included in the actual Hoopes sample of UVLGs (which was based on earlier data releases).  We will use the GALEX data to compare the properties of objects in the MESS to UV selected samples.

\section{SFR Indicators} \label{sfr}

As described in section 1, the luminosity at various wavelengths can be used to estimate SFR.  In this section we draw upon our MIPS data, the SDSS data, and the matches from FIRST, to estimate SFR directly and compare to the B04 total SFR values.  We also check agreement between the various methods.

There is a certain amount of variation in the fixed conversion factors used to relate luminosity and SFR between different authors.  We have chosen to use relations for $SFR_{TIR}$ and $SFR_{1.4GHz}$ from \citet{2003ApJ...586..794B}.  The $SFR_{1.4GHz}$ calibration from \citet{2003ApJ...586..794B} was also applied by \citet{2003ApJ...599..971H} for determining SFR in a large set of star-forming galaxies selected from SDSS.  Both papers assume a Salpeter IMF.  We also use appendix B of \citet{2003ApJ...599..971H} to compute an $SFR_{H\alpha}$ using the emission line fluxes drawn from the MPA/JHU value added catalog.  This formula takes into account a correction from fiber to total $H\alpha$ luminosity, and also an extinction correction determined by the Balmer decrement.  The median $log$ ($L_{H\alpha}$/$L_{\odot}$) obtained for the MESS with this method is $9.2$ .

In order to make a comparison to B04 SFR, we note the conversion for SFR between their Kroupa IMF and the Salpeter IMF is a factor of 1.5.  The conversion factor is the ratio of the mass in the two IMFs for the same amount of ionizing radiation (see section 1 of B04).  The above methods for calculating $SFR_{H\alpha}$, $SFR_{TIR}$, and $SFR_{1.4GHz}$ for the MESS generally indicate a lower SFR than the corresponding values from B04, after taking into account the factor of 1.5.  We calculate median \emph{converted} Kroupa IMF SFRs of 35$M_{\odot}/yr$ for $SFR_{H\alpha}$, and 37$M_{\odot}/yr$ for $SFR_{TIR}$.  Using the flux obtained from the median stacked FIRST cutouts we obtain a luminosity of 9 $\times$ $10^{22}$ $W/Hz$, indicating an $SFR_{1.4GHz}$ of 34 $M_{\odot}/yr$ (again, Kroupa IMF).

Figure \ref{fig9} shows a plot of the $SFR_{1.4GHz}$ versus the B04 total SFR.  The plot indicates essentially no correlation between these measures of SFR for the portion of radio detected objects in our sample.  The sources occupying the highest $SFR_{1.4GHz}$ portion of the diagram do not correspond to either objects with higher positions on the far-IR color-color diagram, or particularly blue galaxies.

Figure \ref{fig10} compares the $SFR_{TIR}$ to $SFR_{1.4GHz}$ which shows reasonable agreement.  The correlation coefficient (Pearson's) between these quantities is $\sim$ $0.8$.  As mentioned previously, the far-IR and radio fluxes do tend to correlate for (U)LIRGs as well as star-forming galaxies in general.  Other authors have observed increased scatter in the Radio-FIR correlation at higher IR luminosities \citep[\eg][]{2003ApJ...586..794B}.  The results of our median stacking of the radio data using 5 bins of increasing IR luminosity are shown as filled squares.  This figure indicates the radio-IR correlation holds for the entire range of MESS redshifts (0.1 $<$ $z$ $<$ 0.3), even though most of these objects were too faint to be detected.  The observed small deviation from linearity at lower luminosities is possibly related to the fact that some of the IR data is based on low SNR measurements.

Figure \ref{fig11} compares $SFR_{TIR}$ to $SFR_{H\alpha}$.  There is increased scatter observed between these two quantities when compared to Fig.~\ref{fig10}.  Considering that the $H\alpha$ luminosity has been corrected for reddening with the Balmer decrement, we do not observe the tight correlation found by \citet{2002AJ....124.3135K} for the Nearby Field Galaxy Survey.  The trend towards the IR direction for higher SFRs also indicates that extinction may not be fully accounted for in some objects.  A similarly weak correlation is found between $SFR_{1.4GHz}$ and $SFR_{H\alpha}$.

Figure \ref{fig12} plots $log$ $L_{TIR}$ versus the specific star formation rate, SFR per unit mass, calculated using the total stellar mass $M_{*}$ (extracted directly from the B04 SFR catalogs), and the $SFR_{TIR}$.  This gives a measure of which galaxies are the most prolific at forming stars relative to their mass.  Since we selected the MESS sample based on high absolute SFR, a consequence was that our galaxies also have high mass.  Nevertheless, our results also indicate the MESS galaxies do have relatively high specific SFR. This plot may show a weak trend toward higher IR luminosities at higher specific SFR.  There is an apparent difference for the UV luminous objects, many of which are not following this trend.

Figure \ref{fig13} plots the $log$ $M_{*}$ versus the oxygen abundance in units of 12$+$ $log$ (O/H) for the sample.  This is otherwise known as the mass-metallicity relation.  The majority of the MESS galaxies fall in a narrow range of high stellar mass objects with $log$ ($M_{*}$/$M_{\odot}$) between 10.5 and 11.5.  In Fig.~ \ref{fig14} we plot the oxygen abundance versus the specific SFR.  The filled squares represent UV luminous galaxies.  Interestingly, the objects with lower specific SFR are the ones with the highest 12$+$ $log$ (O/H).  It is worth emphasizing again that our selection criteria excluded certain types of low continuum luminosity objects which will not be represented here, and may not follow the apparent trend.

\section{Discussion} \label{disc}

\subsection{Composition of the MESS}

The MESS is comprised of galaxies with some of the highest SFRs measured at low redshift, and selected using the SDSS.  Multiple lines of evidence support the conclusion that the MESS are starburst galaxies.  Among them are their IR luminosities, their position on the BPT diagrams, their optical colors, $H\alpha$ luminosities, and the fact that a large percentage are UV and radio detected (given the limits at their redshifts).  As such, the MESS represent a region of parameter space largely unexplored by previous detailed multi-wavelength studies (until quite recently).

Our {\it Spitzer} results have shown that high SFRs determined from optical emission lines via the methods in B04 frequently do correspond to galaxies with high IR luminosities.  The MESS is comprised of 132 LIRGs (although some are in this category based on low SNR data) , five ULIRGs (J082355$+$244830, J110755$+$452809, J120031$+$083114, J140337$+$370355, J142221$+$452011) and one ``IR galaxy'' with $L_{TIR}$ $>$ $10^{8}$ $L_{\odot}$ (J040210$-$054630).  This is a non-trivial result, given both the rarity of objects having LIRG luminosities (in this redshift range), and the fact that we used an optical selection criteria.  We have performed a simple lower limit space density calculation based on the size of the SDSS DR4 spectroscopic footprint, and the corresponding co-moving volume between 0.1 $<$ $z$ $<$ 0.3.  This would indicate a space density for the MESS objects of $\sim$ 2 $\times$ $10^{-7}$ per $Mpc^{3}$.  This is the same estimated space density of classically-selected nearby ULIRGs as mentioned in section \ref{intropart2}, but significantly lower than that of classically-selected LIRGs\footnote{Note that this calculation is based on the DR4 footprint, within which the sample was complete, not the DR7 release}.  In addition to LIRGs, the MESS contains a significant fraction of UVLGs, and spans a large range in physical characteristics, like dust content.

\subsection{Relationship to Other Samples}

The \citet{2007ApJS..173..441H} sample of UVLGs is contained within SDSS, and is included in the MPA/JHU value added catalog, and B04's SFR catalog.  We extract the average of the total SFR distribution determined for these objects from B04's catalog.  The median SFR (Kroupa IMF) for the Hoopes sample is $\sim$ 14$M_{\odot}/yr$, with the highest SFR being 139$M_{\odot}/yr$.  The median $E(B-V)$ is $0.316$.  Overall, the UVLGs tend to be much less dusty and have lower B04 estimated SFRs.  The Hoopes sample is divided up into ``compact'', ``supercompact'', and normal UVLGs based on surface brightness.  As a measure of the surface brightness in the MESS, we have divided the UV luminosities by the area enclosed at the ``ExpRad'' light radius obtained from the SDSS catalogues.  \citet{2007ApJS..173..441H} define the compact UVLGs as having surface brightness $\geq$ $10^{8}$ $L_{\odot}/kpc^2$, and super compact with surface brightness $\geq$ $10^{9}$ $L_{\odot}/kpc^2$.  Seventeen of the 20 UVLGs in the MESS would potentially fall into the compact category, and a handful are possible supercompact UVLGs as well.  However, the uncertainties in determining the radius, due to varying seeing conditions and resolution constraints in SDSS data, prevent a robust measurement of this fraction at this time.

A comparison to classically-selected LIRGs is made difficult by the limited size of the SDSS spectroscopic survey footprint.  Since the {\it IRAS} survey was all-sky, only a fraction of either the GOALS, the FIRST sample, or the 1 Jy Sample are found in the MPA/JHU catalogs.  \citet{2006ApJ...637..227C} report a mean $A_{v}$ of $\sim$ 2.5 for a sample of LIRGs with measured Balmer lines in the {\it Spitzer} First Look Survey, which is comparable to the E(B-V) $\sim$ 0.80 obtained for HII-like ULIRGs in the 1 Jy sample.  In either case, the differences between these and the MESS are not large.  Considering the factor of $\sim$10 increase in IR luminosity for ULIRGs, the increase in the Balmer decrement is not proportionally larger for the dustiest objects.

Taking into account the way the MESS were selected and their overall properties, it appears the MESS represent galaxies bridging a gap between the UVLGs and classically-selected (U)LIRGs, with some overlap on either side.  The majority of them probably suffer from too much extinction to be highly luminous in the far-UV, but are not quite as dusty as the samples of classically-selected (U)LIRGs.  Perhaps a better way to view LIRGs in the MESS would be as a subset of the overall LIRG population -- that is, the population of LIRGs with the least obscured emission line (HII) regions.  This is reflected by the relatively high $L_{H\alpha}$ obtained for the MESS (median $log$ $L_{H\alpha}$/$L_{\odot}$ $=$ 9.2 ).

\citet{2008ApJ...677...37O} has found the starburst activity in some compact UVLGs to be merger/interaction induced based on HST observations.  If these different samples (UVLGs, MESS, classically-selected (U)LIRGs) are indicative of objects at different ages, it would appear that as star formation ramps up, the amount of dust becomes too high for them to be easily detected in the UV.  At even higher SFRs, the optical lines such as $H_{\alpha}$ also become attenuated by dust, leaving the far-IR and radio as the only reliable way of identifying such objects.  This is not to say classically-selected (U)LIRGs do not show strong emission lines, but rather, they are observed at a level underestimating the actual SFRs.  In contrast, the MESS would represent a brief period when the emission lines are still apparent enough to be easily selected.  This scenario could explain the lower space density of the MESS as well.  Thus in this sense, it is not surprising it was difficult to find starburst powered ULIRGs in the B04 SFR catalogs that met our initial criteria.

\subsection{Relationship between Optical Spectra and far-IR}

\begin{figure}
\epsscale{.75}
 \plotone{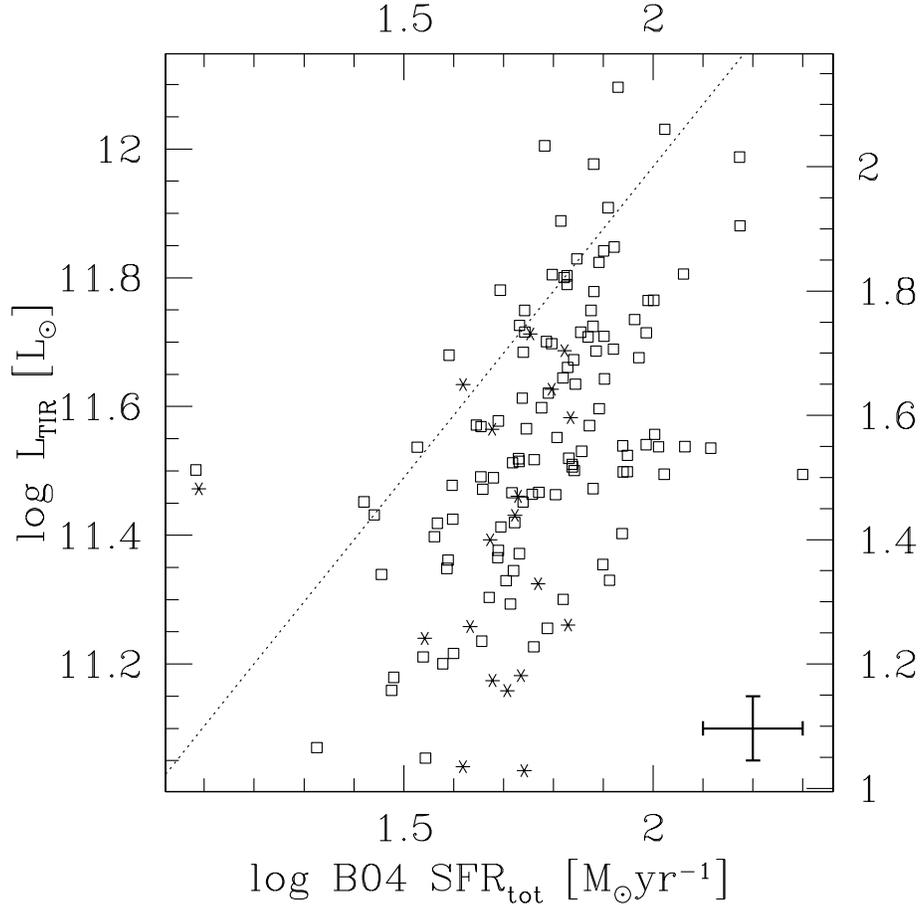}
 \caption{$L_{TIR}$ (left axis) for the MESS versus the DR7 version of the B04 total SFR.  The corresponding $SFR_{TIR}$ is indicated on the right hand axis.  The values for $SFR_{TIR}$ have been converted to their Kroupa IMF equivalents.  Points based on low SNR MIPS measurements are indicated by asterisks.  Typical 1 $\sigma$ error bars shown for B04 SFR are drawn from the 16 and 84 percentiles of the likelihood distributions for SFR.  Error bars shown for $L_{TIR}$ result from uncertainties in the MIPS fluxes themselves, rather than in the SED shape characterization, which will also contribute some uncertainty.  In the cases where values were extracted from the MPA/JHU catalogues we rely on the percentiles when available.  Elsewhere in the paper, typical errors shown are uncertainties in the underlying measurements only, rather than on the direct conversions from luminosities to SFR.  \label{fig5}}
 \end{figure}

\begin{figure}
\epsscale{.75}
 \plotone{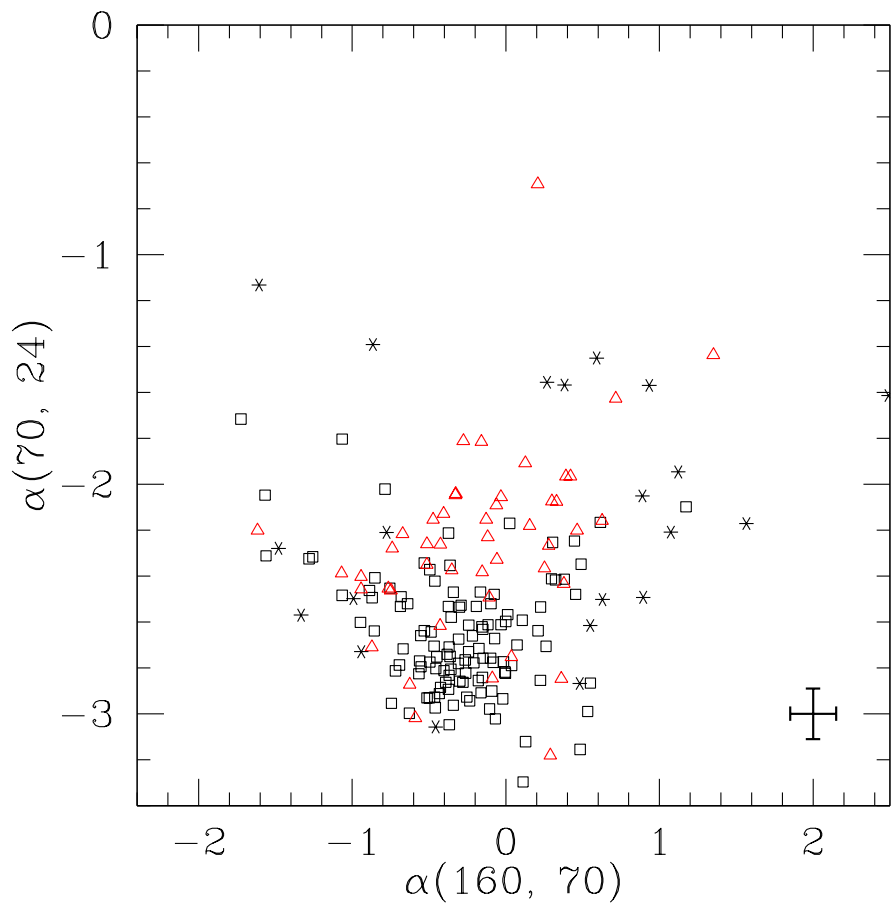}
 \caption{The far-IR color-color diagram (observed frame) adapted to the MIPS bands (originally \citet{1994ApJ...436..102L} and see also \citet{2001ApJ...555..719C}), for the MESS (black squares).  Also plotted are a subset of the GOALS objects \citep{2009PASP..121..559A} (red triangles), for which MIPS fluxes have been released.  This diagram is sometimes used to separate ``warm'' vs ``cold'' (U)LIRGs.  Asterisk symbols indicate low SNR measurements.  \label{fig6}}
 \end{figure}

\begin{figure}
\epsscale{.75}
 \plotone{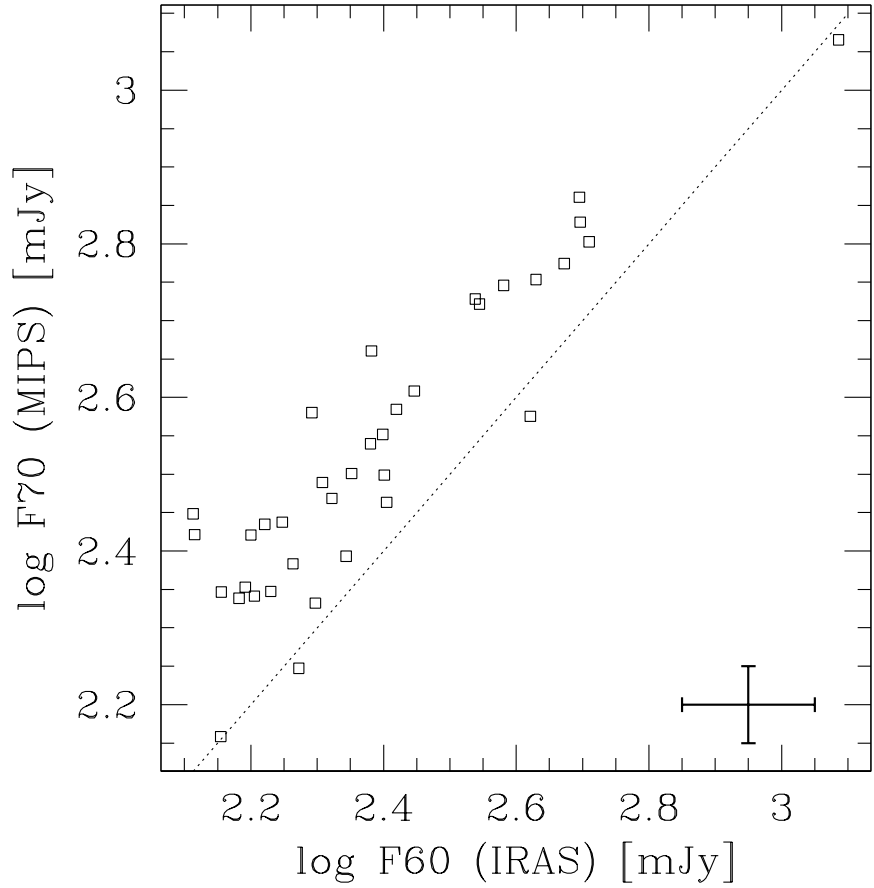}
 \caption{A direct comparison between MIPS 70 $\mu m$ and {\it IRAS} 60 $\mu m$ for all MESS that are found in the {\it IRAS} FSC and FSCR catalogs.  The {\it IRAS} 60 $\mu m$ data are based on moderate and high quality measurements indicated by the Fqual flag. The dotted line represents a one-to-one correspondence.   
\label{fig7}}
 \end{figure}

\begin{figure}
\epsscale{.75}
 \plotone{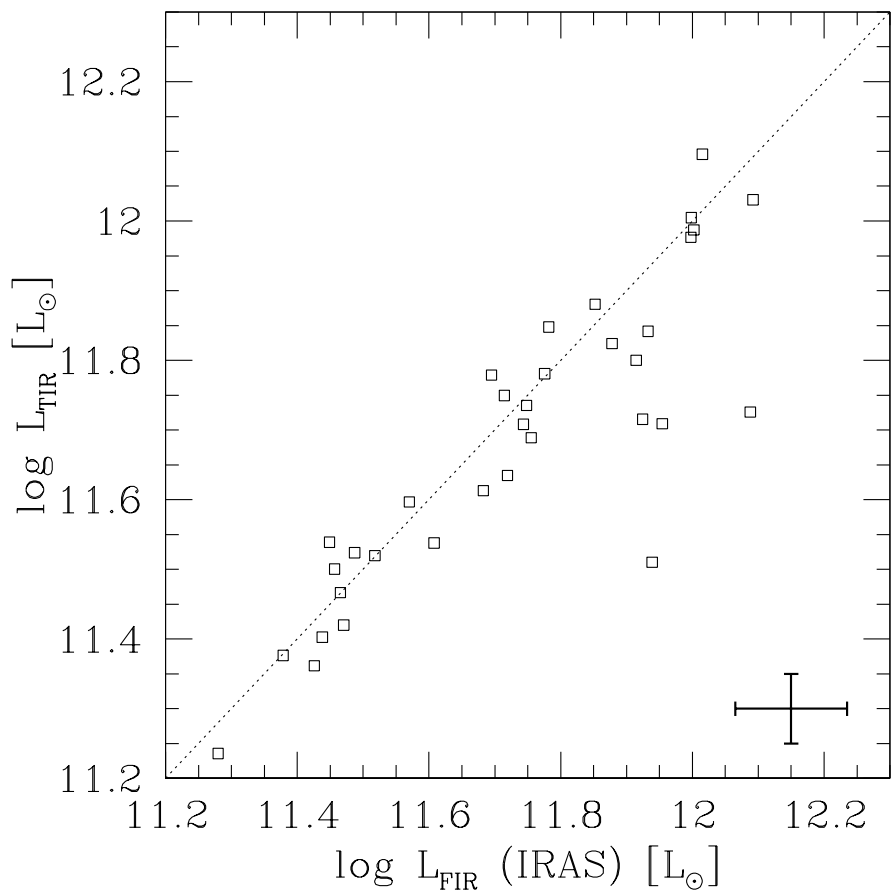}
 \caption{The $L_{TIR}$ computed following methods in \citet{2002ApJ...576..159D} versus the $L_{FIR}$ computed following \citet{1996ARA&A..34..749S} with the {\it IRAS} 60 $\mu m$ (moderate to high quality) and 100 $\mu m$ (mainly upper limits) fluxes.  The values for $L_{FIR}$ are therefore upper limits.  The dotted line represents a one-to-one correspondence.  \label{fig8}}
 \end{figure}

\begin{figure}
\epsscale{.75}
 \plotone{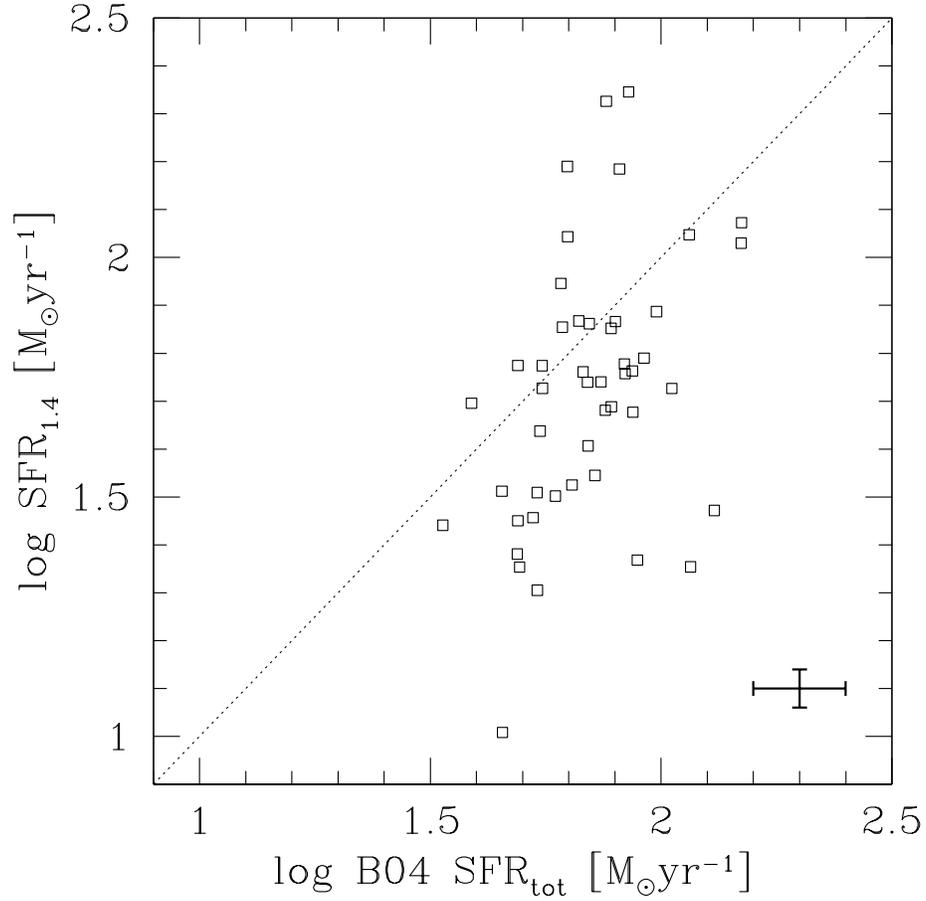}
 \caption{The $SFR_{1.4}$ calculated using radio luminosities from the FIRST survey \citep{1995ApJ...450..559B} and following formulas in \citet{2003ApJ...586..794B} versus the B04 total SFR.  The $SFR_{1.4}$ has been converted from Salpeter to Kroupa IMF.  The dotted line would indicate a one-to-one correspondence, which obviously is not reflected by the data. \label{fig9}}
 \end{figure}

\begin{figure}
\epsscale{.75}
 \plotone{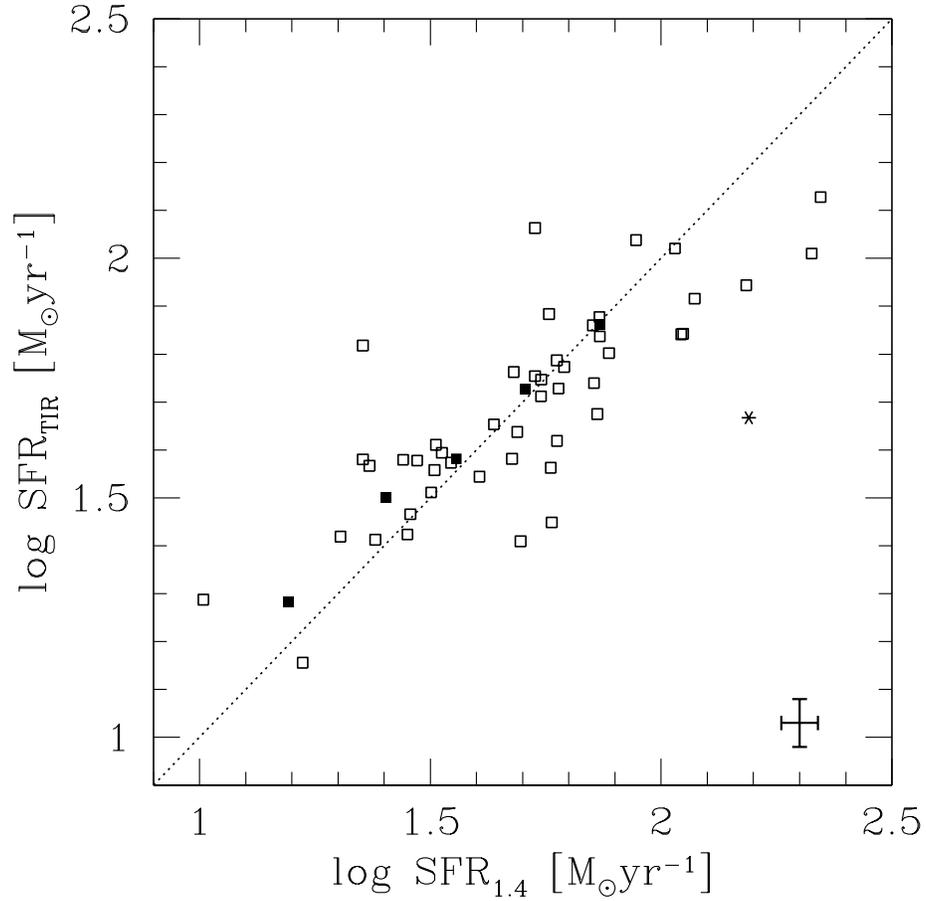}
 \caption{The $SFR_{TIR}$ following \citet{2003ApJ...586..794B} versus $SFR_{1.4}$ (for Kroupa IMF).  Shown in bold are the results of a radio stacking technique applied to all 117 objects within the coverage area of FIRST (see text for details).  The dotted line represents a one-to-one correspondence. \label{fig10}}
 \end{figure}

\begin{figure}
\epsscale{.75}
 \plotone{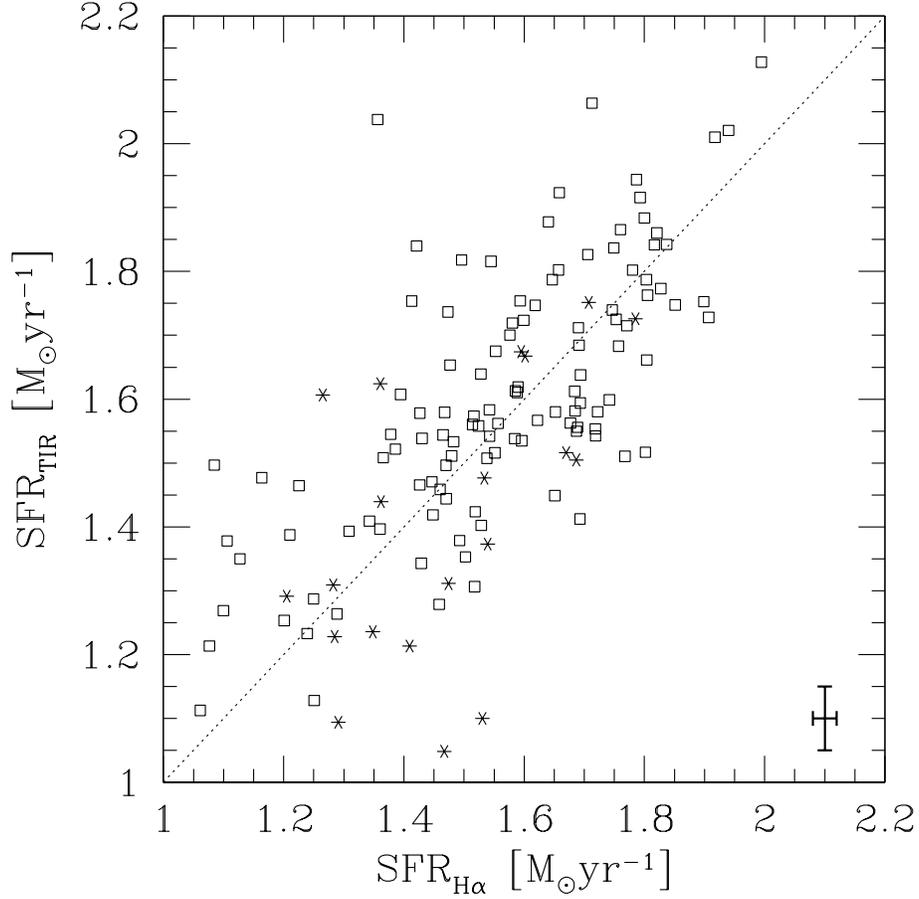}
 \caption{The $SFR_{TIR}$ versus the SFR indicated by the $L_{H\alpha}$ following \citet{2003ApJ...599..971H} and converted to Kroupa IMF.  The appropriate formula accounts for an aperture correction, and dust extinction according to the Balmer decrement.  Asterisk symbols indicate low SNR measurements.  The dotted line represents a one-to-one correspondence. \label{fig11}}
 \end{figure}

\begin{figure}
\epsscale{.75}
 \plotone{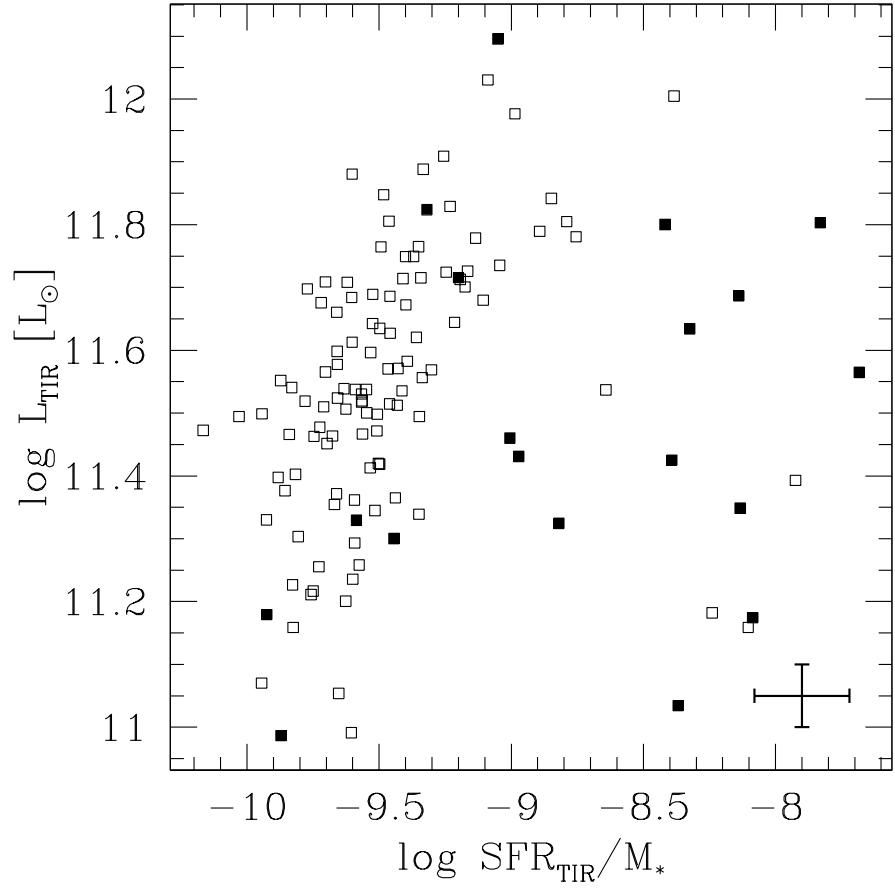}
 \caption{The $log$ $L_{TIR}$ versus $log$ of the specific SFR, calculated from $SFR_{TIR}$ and $M_{*}$ (stellar mass).  Filled squares are UV luminous objects.  \label{fig12}}
 \end{figure}

\begin{figure}
\epsscale{.75}
 \plotone{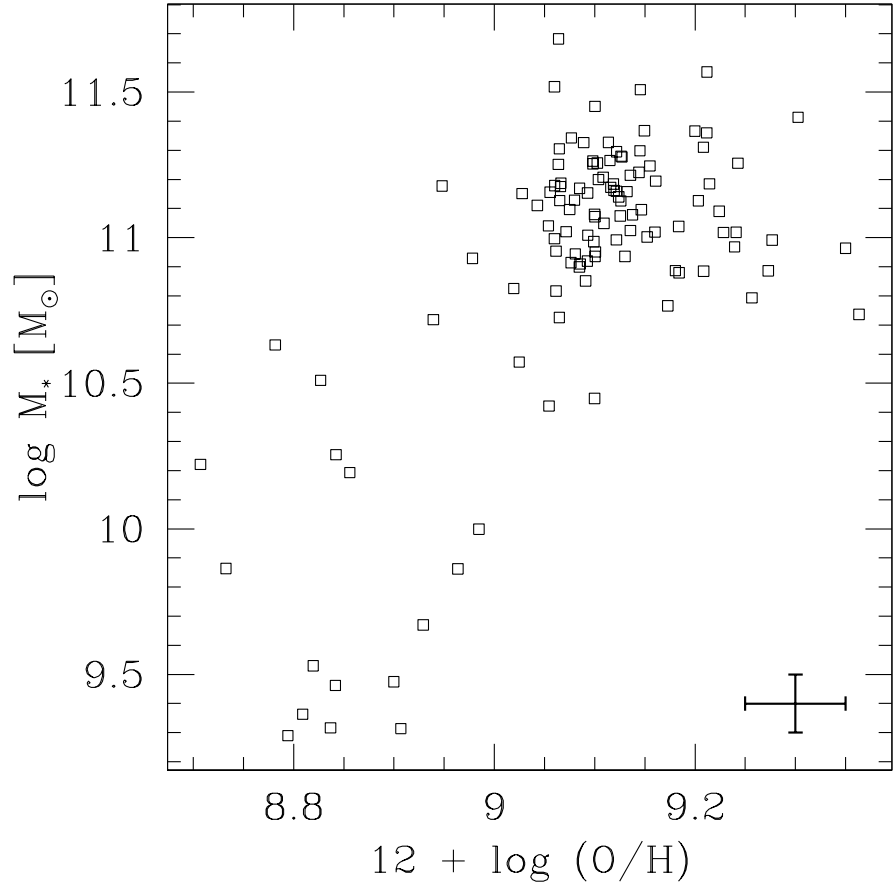}
 \caption{The $log$ $M_{*}$ versus the gas-phase oxygen abundance (metallicity) in units of 12$+$ $log$ (O/H).    \label{fig13}}
 \end{figure}

\begin{figure}
\epsscale{.75}
 \plotone{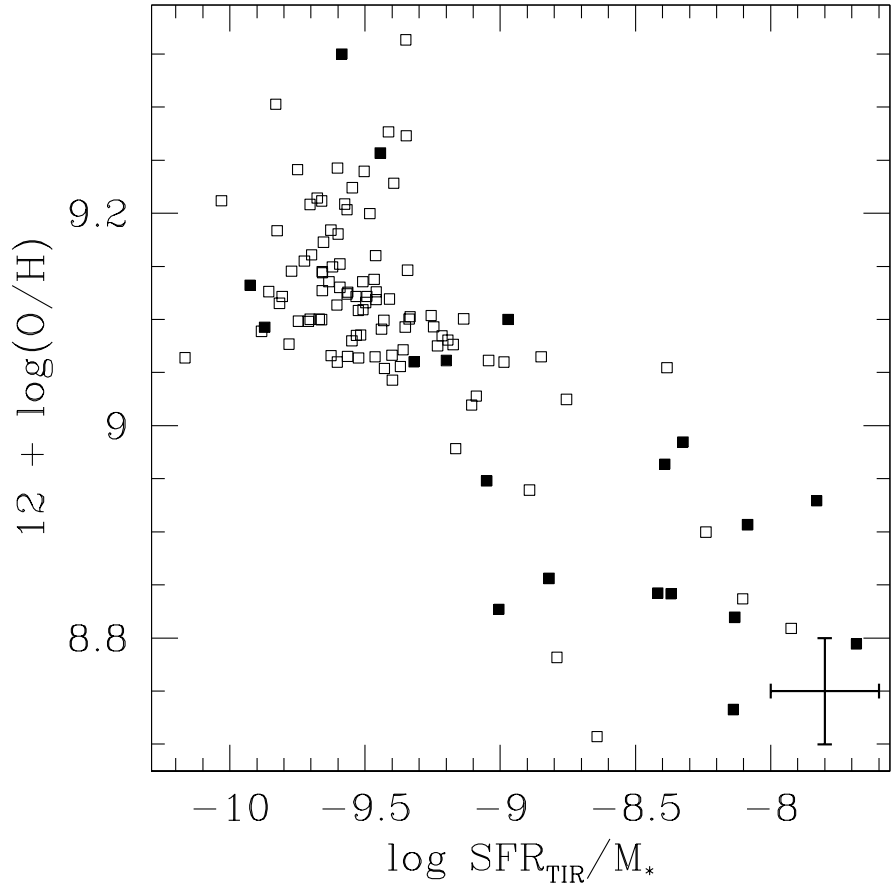}
 \caption{Metallicity in units of 12$+$ $log$ (O/H) versus $log$ of the specific SFR.  Filled squares are UV luminous objects. \label{fig14}}
 \end{figure}

In Fig.~\ref{fig5} we observe a somewhat scattered relation between $L_{TIR}$ and B04 SFR from $log$ $SFR_{tot}$ $=$ 1.50 to 2.1, after which there are too few data points to infer a conclusion.  The right hand axis compares the indicated $SFR_{TIR}$ (Kroupa IMF) to the B04 total SFR.  There is obviously not a good correlation between the IR luminosity and the B04 optical SFR for the MESS.  We have also compared the B04 fiber SFR values to the $SFR_{IR}$, and the plot is also scattered.  It is worth reiterating at this point that B04 SFR is factoring in a correction for extinction, through the spectra model grids.

In an effort to better understand the relationship between B04 SFR and IR luminosity in Fig.~\ref{fig5}, we recall that to first order, the B04 method is based on the strength of the $H\alpha$ line and the Balmer decrement.  For models of emission line HII galaxies, Balmer line strength is strongly affected by the formation of young massive stars.  With these facts in mind, we have a basis to infer that deviations from a direct correspondence between B04 SFR and $L_{H\alpha}$ should be indicative of a higher dust content.

\begin{figure}
\epsscale{.75}
 \plotone{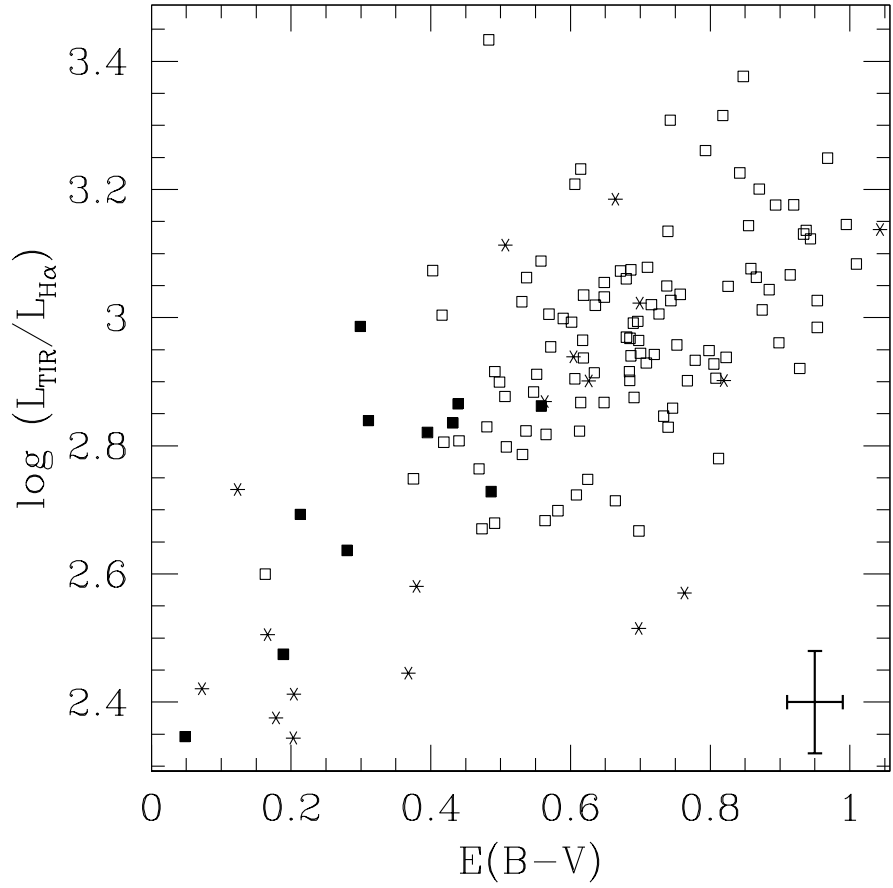}
 \caption{The $log$ $L_{TIR}$/$L_{H\alpha}$ versus the $E(B-V)$ derived from the Balmer decrement.  Filled squares are UV luminous.  Asterisk symbols indicate low SNR measurements.  \label{fig15}}
 \end{figure}

\begin{figure}
\epsscale{.75}
 \plotone{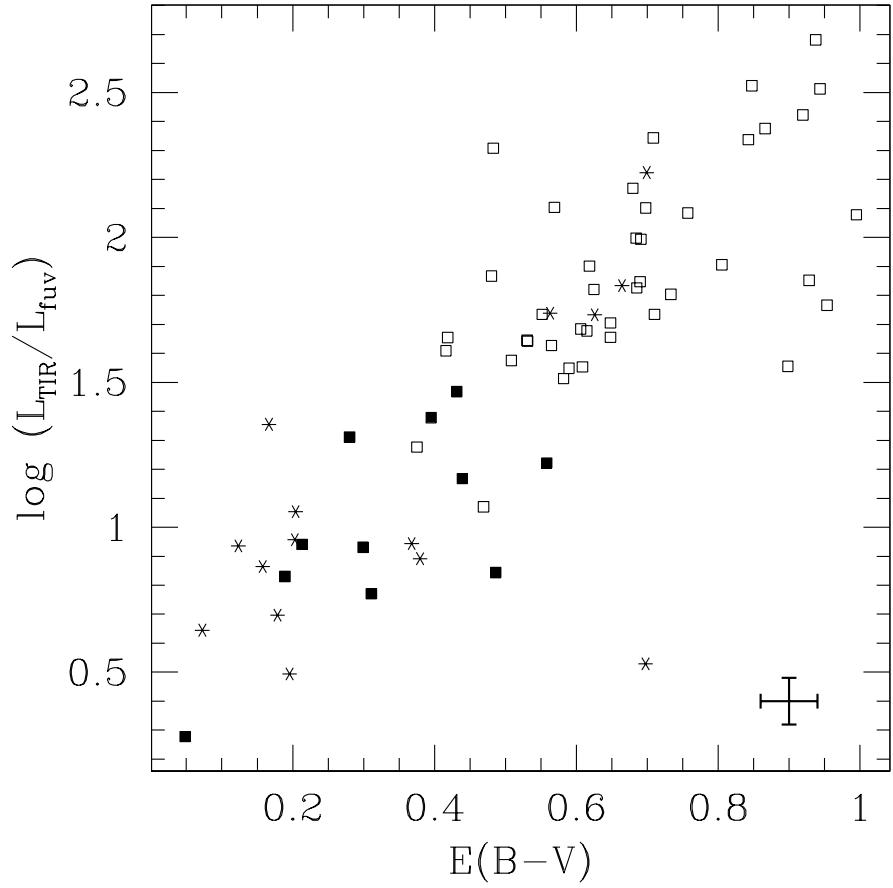}
 \caption{The $log$ $L_{TIR}$/$L_{fuv}$ versus the $E(B-V)$.  Filled squares are UV luminous.  Asterisk symbols indicate low SNR measurements. \label{fig16}}
 \end{figure}

In Fig.~\ref{fig15} we relate the ratio of $L_{TIR}$ to $L_{H\alpha}$ ($H\alpha$ flux not corrected for extinction in this case) compared to the Balmer decrement indicated $E(B-V)$.  This plot indicates the Balmer decrement is correlated with larger $L_{TIR}$/$L_{H\alpha}$ for the MESS.  We have indicated the UVLG MESS by filled squares.  This reflects the fact that dustier galaxies will have larger ``IR excess'' or obscuration by dust of large amounts of star-forming activity.  This relationship is seen in other samples of young star-forming galaxies \citep{2002ApJS..143...47D,2007ApJ...654..226R}.  Similarly, in Fig.~\ref{fig16} we relate the ratio of $L_{TIR}$ to $L_{fuv}$ compared to the Balmer decrement indicated $E(B-V)$, and observe a similar correlation.  Both of these ratios appear to be good measures of the dust content for our sample.  The relationship between the $L_{TIR}$/$L_{fuv}$ fraction and UV spectral slope, and its relationship to dust (sometimes called the IRX-$\beta$ relation) has been documented previously by \citet{1999ApJ...521...64M}, \citet{2005ApJ...619L..51B} and others.  In particular, \citet{2006ApJ...637..242C} studied the correlations between the $L_{TIR}$/$L_{fuv}$ fraction and the UV spectral slope, with a sample of optically selected normal star-forming galaxies and a separate sample of starbursts.  They observe relationships between the Balmer decrement derived $H_{\alpha}$ attenuation, the UV spectral slope and $L_{TIR}$/$L_{fuv}$ fraction allowing them to estimate this fraction without far-IR data.

\begin{figure}
\epsscale{.75}
\plotone{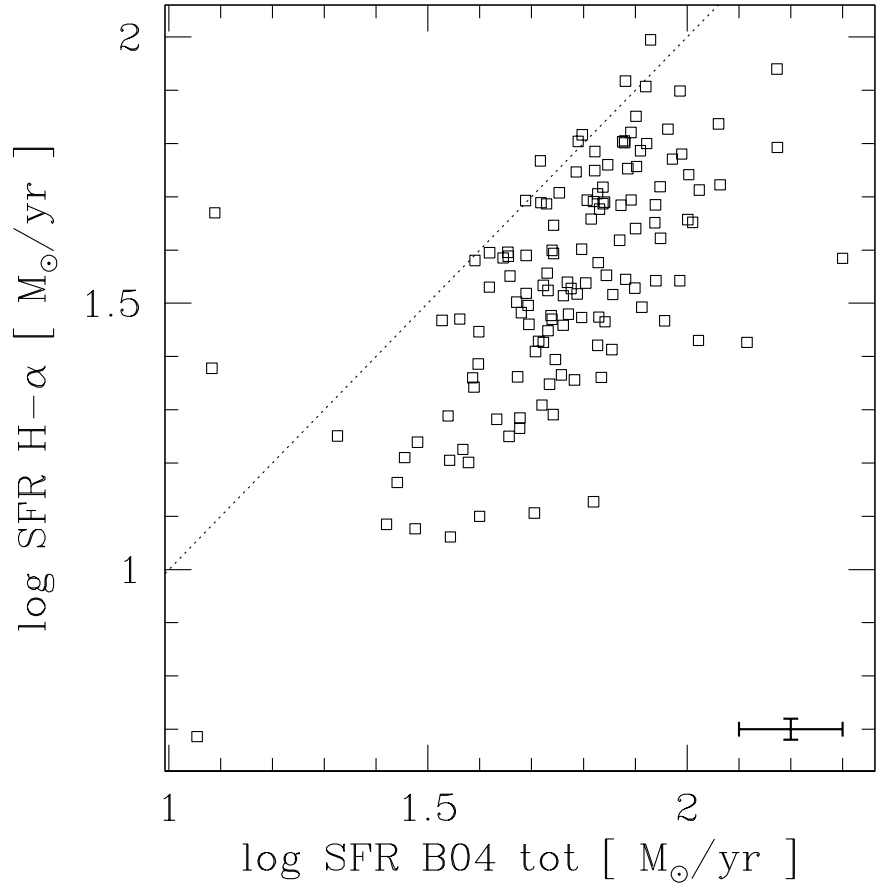}
\caption{The $log$ $SFR_{H\alpha}$ versus the total $log$ B04 $SFR$.  The dotted line represents a one-to-one correspondence.  As mentioned earlier, the representative typical error on the $SFR_{H\alpha}$ is based on the flux measurement uncertainty, and does not include the inherent large uncertainty in going directly from this flux to an SFR.  \label{fig17}}
\end{figure}

In Fig.~\ref{fig11} we compared $SFR_{TIR}$ to the SFR based on $L_{H\alpha}$, corrected via the Balmer decrement.  There is significant scatter in that figure, but it does appear to align along the one-to-one correspondence line.  In Fig.~\ref{fig5}, there is an offset to higher values of B04 SFR for a given $SFR_{TIR}$.   In Fig.~\ref{fig17} we plot the $SFR_{H\alpha}$ versus the B04 total SFR and note a similar deviation from the one-to-one line.   For the MESS sample, the B04 method appears to be assigning higher SFRs than would be indicated by Balmer decrement corrected H-alpha alone.  
Recalling that radio is thought to be unaffected by dust, and taking into account the relatively good agreement in Fig.~\ref{fig10} between $SFR_{1.4GHz}$ and $SFR_{TIR}$, all indications are that there is not a significantly obscured source of star formation that has been missed by the direct methods.

Our reanalysis using the SDSS DR7 values for SFR based on the method of B04 shows that the original selection (based on SDSS DR4) somewhat overestimated the optical SFRs. Even the DR7 values seem to over-predict SFR, by around 50\% at the high SFR end of our sample to 100\% at the low end (see Fig.~\ref{fig5} and Fig.~\ref{fig17}); although, the SFRs as measured by all the multi-wavelength indicators are almost always high enough to put our galaxies in the (U)LIRG range. Using the DR7 values, despite these systematic offsets (perhaps due to an overestimation of the amount of dust attenuation for these objects by the B04 method), the correlation with the other indicators is nevertheless quite clear. The B04 relationships were calibrated to work best for typical low redshift galaxies, and by selecting the most extreme objects from B04, we may be biased to selecting objects where the B04 method overestimates the SFR, which perhaps explains the remaining discrepancies.

Our results have important consequences for high redshift surveys of galaxies relying on emission line fitting methods alone.  It appears the method may not be sufficient, or at least not well calibrated for some of the high SFR galaxies like the MESS.  Note that this problem is significant since starburst galaxies will represent an increasing fraction of the populations as one moves to higher redshifts.  Our next step is to see if there are morphological effects contributing to the scatter.  High SFR galaxies are frequently irregular, and the method for deriving the SFR from the SDSS fiber is likely to be sensitive to this at some level as well.  Additionally, this information should help us sort out a detailed evolutionary scenario for the MESS.  As more information becomes available on samples such as a the $BzK$ selected galaxies it will be useful to compare to those as well.

\newpage

\section{Conclusions} \label{conclusions}

\begin{itemize}
 \item The MESS is composed of galaxies with some of the highest optically determined SFRs measured in the SDSS, and as such probes a region of parameter space not well explored by previous studies attempting to relate an optical SFR, with dust level, and far-IR properties.  
\item Objects with very high optically determined SFRs (sample median is 61 $M_{\odot}$ $yr^{-1}$), as measured by B04's methods for SDSS DR7, often have LIRG level luminosities in the IR.  
\item Previous studies have attempted to find a direct conversion factor between $H\alpha$ luminosity and SFR.  We find that after correcting for extinction, the indicated $SFR_{H\alpha}$ for the MESS is correlated, and roughly in agreement with $SFR_{TIR}$.  Similarly, the 1.4 $GHz$ radio SFRs are also in reasonable agreement with the $SFR_{TIR}$.  Quantitatively, we find our indicated SFRs by these direct methods to be lower than the B04 predictions (after taking into account a conversion between Salpeter and Kroupa IMF).
\item Varying levels of dust extinction are spanned by the MESS, from virtually none to a Balmer decrement indicated $E(B-V)$ of 1.04.  
\item 20 of the MESS objects are found to be UV luminous galaxies, with some being possible ``supercompact'' UVLGs.
\item A correlation is found between $L_{TIR}$/$L_{H\alpha}$ (IR excess), $L_{TIR}$/$L_{fuv}$, and the Balmer decrement.  This relationship has been observed by other authors, and is commonly seen in young dusty starburst galaxies.
\item Based on the above properties we believe the MESS represent a category of luminous starburst galaxies bridging a gap between UVLGs and classically-selected LIRGs. 
\item In a future paper we will examine the near-IR morphologies of galaxies in the MESS sample to see if they are also morphologically intermediate between UVLGs and classically-selected LIRGs.
\end{itemize}

\newpage

\section{Acknowledgements}

\acknowledgments

{\it Facilities:} \facility{IRAS ()}, \facility{Spitzer (MIPS)}, \facility{Sloan ()}, \facility{VLA ()}, \facility{GALEX ()}

We gratefully acknowledge Wim de Vries, Wil van Breugel, and Bob Becker who were co-authors on the original proposal.  We thank Lee Armus, Jarle Brinchmann, Roderik Overzier, and Antara Basu-Zych for their helpful input and responses to our questions.  We thank Daniel Dale for sharing coefficients for IR luminosity and answering questions.    We thank Roser Pello for providing the latest version of Hyperz.  We also thank the team that supports the SSC helpdesk.  We used Edward L. Wright's online and Python scripted Cosmo-Calculator \citep{2006PASP..118.1711W}, and the TOPCAT software for graphical viewing of FITS tables: \emph{http://www.star.bristol.ac.uk/~mbt/topcat/}.  Finally we thank the anonymous referee for their thoughtful insights which greatly improved the paper.  This work is based in part on observations made with the Spitzer Space Telescope (Program ID 40640), which is operated by the Jet Propulsion Laboratory, California Institute of Technology under a contract with NASA. Support for this work was provided by NASA through an award issued by JPL/Caltech.  A portion of this work was also supported at The Aerospace Corporation by the Independent Research and Development program.

\newpage

\appendix
\section{Changes from DR4 to DR7}

\begin{figure}
\epsscale{.75}
\plotone{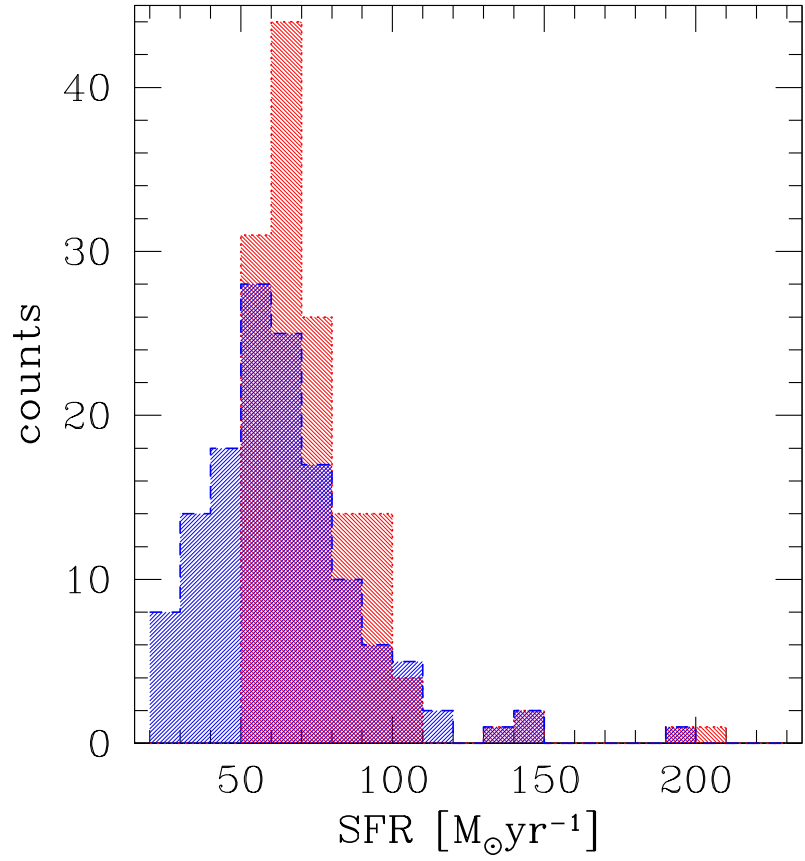}
\caption{Histograms of the MESS B04 total SFR from the DR4 (dotted red line) and DR7 (dashed blue line) versions of the catalog. \label{fig18}}
\end{figure}

\begin{figure}
\epsscale{.75}
\plotone{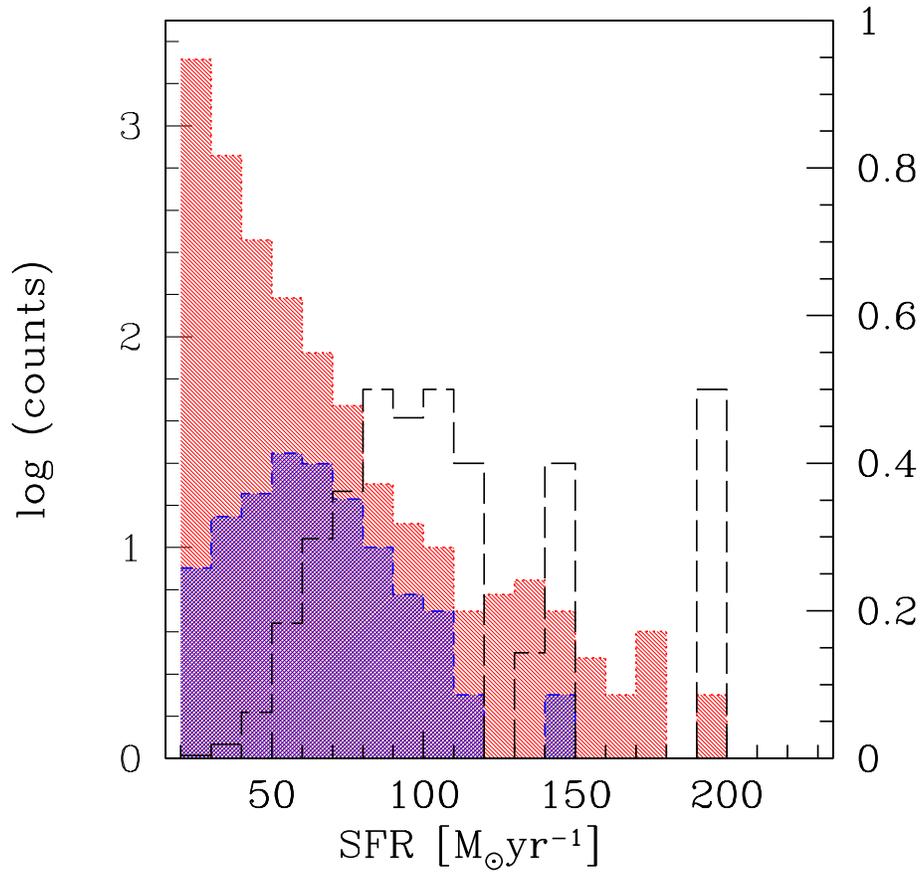}
\caption{Histograms of the DR7 B04 total SFR for the complete sample of star forming galaxies (dotted red line),  and the MESS (dashed blue line), with counts plotted on a log scale (left side).  Plotted in black is the ratio of the two histograms, a measure of the completeness, on a scale from 0 to 1 (right side).  \label{fig19}}
\end{figure}

Shortly after the observing campaign for the MESS was completed, a new version of the B04 catalog was released.  Figure \ref{fig18} compares the B04 total SFR distributions for the MESS in the DR4 and DR7 versions of the sample.  Clearly, the distribution becomes wider, encompassing SFRs that are below 50, but the shift in the peak of the distribution is not large.  Thus the sample selection in DR7 is not much different than we originally intended.  One of our main tasks was then to understand how well the MESS represented the revised catalog.  Given the increased number of sources from the larger footprint, and the changes to the methodology, the MESS can no longer be considered a complete sample.  Shown in Fig.~\ref{fig19} is the total SFR distribution of all unique DR7 galaxies with SFR $\geq$ 20 $M_{\odot}$ $yr^{-1}$ compared to the MESS, with counts on a log scale.  We selected only star formation powered DR7 objects (class $=$ 1), in the same redshift range as the MESS (0.1 $<$ $z$ $<$ 0.3).  Also plotted is the ratio of the two histograms, which can be used as a gauge for completeness.  Between SFR $=$ 20, about the minimum LIRG level, and 50 $M_{\odot}$ $yr^{-1}$, the MESS is only a small fraction of the galaxies in DR7.  However, above 50 $M_{\odot}$ $yr^{-1}$, the MESS completeness quickly increases and levels off at about 50\%.  This makes it reasonably representative of the more extreme SFRs.  In the original DR4 version of the catalog, many of the very high SFRs were spurious, however, improvements in the methodology mean there is reason to believe these new sources are real.  In the future, they should probably be included in an updated version of the MESS.

\bibliographystyle{apj}
\bibliography{messref}

\clearpage

\newpage

\clearpage

\clearpage

\clearpage

\clearpage

\clearpage

\clearpage

\clearpage

 \clearpage

\clearpage

\clearpage

\clearpage

\clearpage

\clearpage

\clearpage

\clearpage

\clearpage

\clearpage

\clearpage

\clearpage

\begin{deluxetable}{lccccc}
\tablecaption{The Sample\tablenotemark{*} \label{tab:tableone}}
\tablecolumns{5}
\tablewidth{0pt}
\tablehead{\colhead{Name}  & \colhead{R.A. (J2000)} & \colhead{Dec. (J2000)}   & \colhead{$z$} & \colhead{$log$ SFR}  \\
\colhead{SDSS}  & \colhead{[$deg$]} & \colhead{[$deg$]}   & \colhead{} & \colhead{[$M_{\odot}/yr$]}  \\
\colhead{(1)} & \colhead{(2)} & \colhead{(3)} & \colhead{(4)}   & \colhead{(5)}  }
\startdata
J001629$-$103511 & 4.1215 & -10.5866 & 0.212 & 1.72 \\
J002334$+$145815 & 5.8935 & 14.9709 & 0.153 & 1.73 \\
J002353$+$155947 & 5.9749 & 15.9966 & 0.192 & 1.75 \\
J003816$-$010911 & 9.5673 & -1.1532 & 0.296 & 2.00 \\
J004236$+$160202 & 10.6539 & 16.0341 & 0.247 & 1.67 \\
J004646$+$154339 & 11.6947 & 15.7277 & 0.181 & 1.48 \\
J005546$+$155603 & 13.9427 & 15.9342 & 0.192 & 1.72 \\
J011101$+$000403 & 17.7548 & 0.0676 & 0.296 & 1.44 \\
J011615$+$144646 & 19.0635 & 14.7796 & 0.180 & 1.33 \\
J012727$-$085943 & 21.8655 & -8.9955 & 0.210 & 1.66 \\
J014547$+$011348 & 26.4473 & 1.2301 & 0.181 & 1.53 \\
J015400$-$081718 & 28.5007 & -8.2884 & 0.166 & 1.69 \\
J020038$-$005954 & 30.1611 & -0.9985 & 0.253 & 1.62 \\
J020215$+$131749 & 30.5657 & 13.2971 & 0.207 & 1.72 \\
J021601$-$010312 & 34.0074 & -1.0534 & 0.289 & 1.63 \\
J022229$+$002900 & 35.6245 & 0.4835 & 0.300 & 1.62 \\
J024750$+$004718 & 41.9592 & 0.7884 & 0.252 & 1.85 \\
J025220$-$004343 & 43.0858 & -0.7287 & 0.298 & 1.42 \\
J025958$-$003622 & 44.9928 & -0.6061 & 0.175 & 1.59 \\
J031036$+$000817 & 47.6541 & 0.1383 & 0.233 & 1.08 \\
J031345$-$010517 & 48.4391 & -1.0883 & 0.257 & 1.57 \\
J032641$+$004847 & 51.6740 & 0.8132 & 0.285 & 1.83 \\
J033206$+$011048 & 53.0279 & 1.1800 & 0.271 & 1.09 \\
J033918$-$011424 & 54.8261 & -1.2402 & 0.270 & 1.54 \\
J034742$+$010959 & 56.9289 & 1.1665 & 0.240 & 1.65 \\
J034830$-$064230 & 57.1260 & -6.7085 & 0.166 & 1.84 \\
J040210$-$054630 & 60.5454 & -5.7751 & 0.270 & 1.96 \\
J073219$+$380508 & 113.0827 & 38.0856 & 0.179 & 1.94 \\
J074936$+$333716 & 117.4032 & 33.6212 & 0.273 & 1.82 \\
J075536$+$250846 & 118.9038 & 25.1462 & 0.239 & 1.80 \\
J080522$+$270829 & 121.3428 & 27.1416 & 0.140 & 1.96 \\
J081841$+$463505 & 124.6750 & 46.5850 & 0.218 & 1.89 \\
J082140$+$032147 & 125.4184 & 3.3632 & 0.192 & 1.92 \\
J082355$+$244830 & 125.9807 & 24.8084 & 0.234 & 2.02 \\
J084800$+$061837 & 132.0038 & 6.3103 & 0.220 & 1.56 \\
J084827$+$331643 & 132.1130 & 33.2787 & 0.109 & 1.66 \\
J085906$+$542150 & 134.7765 & 54.3639 & 0.182 & 1.05 \\
J090244$+$343000 & 135.6836 & 34.5000 & 0.196 & 1.80 \\
J090250$+$334901 & 135.7086 & 33.8171 & 0.116 & 1.89 \\
J090442$+$453317 & 136.1766 & 45.5548 & 0.181 & 1.76 \\
J090949$+$014847 & 137.4567 & 1.8132 & 0.182 & 1.92 \\
J091426$+$102409 & 138.6093 & 10.4027 & 0.176 & 1.81 \\
J092322$+$324830 & 140.8437 & 32.8085 & 0.140 & 1.73 \\
J092456$+$001829 & 141.2350 & 0.3082 & 0.153 & 1.88 \\
J092710$+$010232 & 141.7953 & 1.0423 & 0.169 & 1.90 \\
J092905$+$494059 & 142.2709 & 49.6832 & 0.189 & 1.84 \\
J093613$+$620905 & 144.0572 & 62.1515 & 0.225 & 1.90 \\
J093714$+$120019 & 144.3114 & 12.0055 & 0.140 & 1.73 \\
J094849$-$005314 & 147.2049 & -0.8874 & 0.231 & 1.99 \\
J095618$+$430727 & 149.0763 & 43.1244 & 0.276 & 1.71 \\
J100950$+$552336 & 152.4592 & 55.3935 & 0.194 & 1.78 \\
J101508$+$365818 & 153.7855 & 36.9718 & 0.208 & 1.76 \\
J101636$-$011358 & 154.1534 & -1.2329 & 0.172 & 1.84 \\
J101732$+$140436 & 154.3863 & 14.0769 & 0.231 & 1.91 \\
J102822$+$405558 & 157.0919 & 40.9328 & 0.203 & 1.87 \\
J102944$+$525143 & 157.4373 & 52.8622 & 0.227 & 1.79 \\
J104116$+$565345 & 160.3167 & 56.8959 & 0.185 & 1.83 \\
J104729$+$572842 & 161.8744 & 57.4786 & 0.230 & 1.80 \\
J104906$+$015920 & 162.2782 & 1.9889 & 0.227 & 1.94 \\
J105527$+$064015 & 163.8633 & 6.6708 & 0.173 & 1.79 \\
J110618$+$582441 & 166.5779 & 58.4116 & 0.125 & 1.71 \\
J110755$+$452809 & 166.9828 & 45.4694 & 0.272 & 1.93 \\
J110908$+$534143 & 167.2856 & 53.6955 & 0.199 & 1.69 \\
J111929$+$011117 & 169.8724 & 1.1882 & 0.185 & 2.01 \\
J112152$+$414757 & 170.4684 & 41.7994 & 0.195 & 1.74 \\
J112436$+$054053 & 171.1525 & 5.6815 & 0.233 & 1.85 \\
J112851$+$413455 & 172.2158 & 41.5822 & 0.181 & 1.67 \\
J113513$+$470821 & 173.8075 & 47.1392 & 0.130 & 1.72 \\
J113703$+$504420 & 174.2655 & 50.7391 & 0.160 & 1.69 \\
J115111$+$104710\tablenotemark{\dagger} & 177.7967 & 10.7862 & 0.115 & 1.84 \\
J115630$+$500822 & 179.1276 & 50.1395 & 0.236 & 1.59 \\
J115744$+$120750 & 179.4348 & 12.1308 & 0.183 & 1.74 \\
J120031$+$083114 & 180.1307 & 8.5207 & 0.248 & 2.17 \\
J120204$+$495112 & 180.5190 & 49.8534 & 0.287 & 1.83 \\
J120805$+$542258 & 182.0246 & 54.3830 & 0.286 & 1.46 \\
J121005$+$002640 & 182.5248 & 0.4445 & 0.128 & 1.69 \\
J122016$+$534028 & 185.0667 & 53.6747 & 0.197 & 1.88 \\
J122320$+$115931 & 185.8341 & 11.9921 & 0.165 & 2.12 \\
J122641$-$000620 & 186.6737 & -0.1058 & 0.279 & 1.80 \\
J123117$+$015430 & 187.8242 & 1.9084 & 0.269 & 1.71 \\
J123552$+$592400 & 188.9679 & 59.4003 & 0.178 & 2.06 \\
J123645$+$535901 & 189.1902 & 53.9838 & 0.195 & 1.88 \\
J124137$+$444453 & 190.4052 & 44.7483 & 0.192 & 1.82 \\
J124907$+$582729 & 192.2804 & 58.4581 & 0.297 & 1.77 \\
J125045$+$490640 & 192.6895 & 49.1113 & 0.218 & 1.91 \\
J125410$+$035951 & 193.5435 & 3.9976 & 0.168 & 1.74 \\
J125548$+$505716 & 193.9518 & 50.9546 & 0.151 & 1.79 \\
J130553$+$110319 & 196.4729 & 11.0554 & 0.238 & 1.60 \\
J130704$+$485845 & 196.7689 & 48.9793 & 0.123 & 1.95 \\
J130847$+$504259 & 197.1987 & 50.7166 & 0.124 & 1.59 \\
J130919$+$055049 & 197.3324 & 5.8469 & 0.274 & 1.74 \\
J131101$-$004215 & 197.7573 & -0.7042 & 0.245 & 1.73 \\
J131447$+$012759 & 198.6965 & 1.4665 & 0.287 & 1.82 \\
J131810$+$041929 & 199.5426 & 4.3247 & 0.113 & 2.17 \\
J132034$+$443649 & 200.1434 & 44.6138 & 0.166 & 1.86 \\
J133114$+$583342 & 202.8114 & 58.5617 & 0.196 & 1.89 \\
J134619$+$115204 & 206.5819 & 11.8680 & 0.195 & 2.30 \\
J134911$+$021716 & 207.2963 & 2.2878 & 0.218 & 2.00 \\
J135435$-$012213 & 208.6467 & -1.3705 & 0.134 & 1.84 \\
J135646$+$465414 & 209.1933 & 46.9041 & 0.214 & 1.99 \\
J140337$+$370355 & 210.9073 & 37.0654 & 0.211 & 1.78 \\
J141803$+$534104 & 214.5150 & 53.6845 & 0.164 & 1.83 \\
J142057$+$015232 & 215.2416 & 1.8756 & 0.265 & 1.90 \\
J142221$+$452011 & 215.5910 & 45.3366 & 0.167 & 1.88 \\
J143047$+$032330 & 217.6984 & 3.3917 & 0.167 & 1.54 \\
J143727$+$394530 & 219.3633 & 39.7586 & 0.180 & 1.60 \\
J145435$+$452856 & 223.6482 & 45.4823 & 0.269 & 1.68 \\
J150627$+$562702 & 226.6166 & 56.4507 & 0.279 & 1.73 \\
J150705$+$610919 & 226.7739 & 61.1555 & 0.183 & 1.84 \\
J151226$+$462903 & 228.1089 & 46.4844 & 0.205 & 1.76 \\
J151320$-$002551 & 228.3359 & -0.4311 & 0.218 & 1.74 \\
J151405$+$432528 & 228.5228 & 43.4246 & 0.208 & 1.83 \\
J152044$+$321440 & 230.1874 & 32.2446 & 0.132 & 1.94 \\
J152552$+$041732 & 231.4691 & 4.2924 & 0.198 & 1.54 \\
J153428$+$315314 & 233.6205 & 31.8874 & 0.109 & 1.77 \\
J154049$+$390350 & 235.2051 & 39.0641 & 0.239 & 1.82 \\
J154120$+$453619 & 235.3334 & 45.6053 & 0.203 & 1.68 \\
J154652$+$030402 & 236.7205 & 3.0674 & 0.165 & 1.66 \\
J155707$+$050530 & 239.2792 & 5.0918 & 0.139 & 1.58 \\
J155934$+$404144 & 239.8946 & 40.6956 & 0.298 & 2.02 \\
J160531$+$401741 & 241.3813 & 40.2948 & 0.244 & 1.88 \\
J161210$-$005756 & 243.0417 & -0.9657 & 0.218 & 1.73 \\
J161401$+$423721 & 243.5056 & 42.6228 & 0.137 & 1.69 \\
J163216$+$352449 & 248.0678 & 35.4138 & 0.255 & 1.83 \\
J204719$-$004931 & 311.8297 & -0.8255 & 0.156 & 1.90 \\
J205013$-$011521 & 312.5561 & -1.2559 & 0.256 & 1.72 \\
J205308$+$010937 & 313.2869 & 1.1605 & 0.172 & 1.48 \\
J210256$+$000955 & 315.7341 & 0.1655 & 0.191 & 1.87 \\
J210420$-$061840 & 316.0854 & -6.3114 & 0.271 & 1.99 \\
J211729$-$000410 & 319.3733 & -0.0696 & 0.210 & 1.82 \\
J213822$+$105132 & 324.5941 & 10.8591 & 0.219 & 1.95 \\
J213951$-$082538 & 324.9638 & -8.4274 & 0.155 & 1.74 \\
J221950$+$000125 & 334.9607 & 0.0237 & 0.231 & 1.60 \\
J222100$-$002537 & 335.2508 & -0.4272 & 0.198 & 1.68 \\
J223528$+$135812 & 338.8693 & 13.9702 & 0.183 & 1.97 \\
J233417$+$010353 & 353.5717 & 1.0649 & 0.281 & 1.65 \\
J234143$-$094048 & 355.4319 & -9.6800 & 0.275 & 2.06 \\
J235237$-$102943 & 358.1577 & -10.4955 & 0.250 & 1.75 \\
\enddata
\tablenotetext{*}{Columns: (1) SDSS name. (2) J2000 RA. (3) J2000 DEC. (4) Spectroscopic Redshift from SDSS. (5) log B04  DR7 total SFR. }
\tablenotetext{\dagger}{This source was not included in the DR7 SFR release, so the SFR is from the DR4 release.}
\end{deluxetable}

\clearpage

\begin{deluxetable}{lcccccccc}
\tablecaption{MIPS Photometry\tablenotemark{*}  \label{tab:tabletwo} }
\tablecolumns{8}
\tablehead{\colhead{Name}  & \colhead{$f_{\nu}$(24$\mu m$)} & \colhead{$\Delta$(24)}   & \colhead{$f_{\nu}$(70$\mu m$)} & \colhead{$\Delta$(70)} & \colhead{$f_{\nu}$(160$\mu m$)} & \colhead{$\Delta$(160)} & \colhead{$log$ $L_{TIR}$} \\
\colhead{SDSS}  & \colhead{[$mJy$]} & \colhead{[$mJy$]} & \colhead{[$mJy$]} & \colhead{[$mJy$]} & \colhead{[$mJy$]} & \colhead{[$mJy$]} & \colhead{$L_{\odot}$} \\
\colhead{(1)} & \colhead{(2)} & \colhead{(3)} & \colhead{(4)}   & \colhead{(5)} & \colhead{(6)} & \colhead{(7)} & \colhead{(8)} }
\startdata
J001629$-$103511 & 7.9 & 0.4 & 111.5 & 11.5 & 197.7 & 29.8 & 11.5 \\
J002334$+$145815 & 12.4 & 0.6 & 283.4 & 28.6 & 392.7 & 59.0 & 11.5 \\
J002353$+$155947 & 7.8 & 0.4 & 195.0 & 19.8 & 276.2 & 41.6 & 11.6 \\
J003816$-$010911 & 6.7 & 0.4 & 117.0 & 12.1 & 141.3 & 21.3 & 11.8 \\
J004236$+$160202 & 10.8 & 0.6 & 55.4 & 6.0 & 26.8 & 22.0 & 11.4 \\
J004646$+$154339 & 5.1 & 0.3 & 43.3 & 5.0 & 165.1 & 25.0 & 11.2 \\
J005546$+$155603 & 7.0 & 0.4 & 137.6 & 14.2 & 229.6 & 34.7 & 11.5 \\
J011101$+$000403 & 2.1 & 0.2 & 68.5 & 7.4 & 65.2 & 10.1 & 11.4 \\
J011615$+$144646 & 5.7 & 0.3 & 47.6 & 5.5 & 95.3 & 14.7 & 11.1 \\
J012727$-$085943 & 5.8 & 0.3 & 122.0 & 12.5 & 181.3 & 27.4 & 11.5 \\
J014547$+$011348 & 18.9 & 1.0 & 223.6 & 22.6 & 155.8 & 23.5 & 11.5 \\
J015400$-$081718 & 31.7 & 1.6 & 376.1 & 37.8 & 530.6 & 79.7 & 11.8 \\
J020038$-$005954 & 16.7 & 0.8 & 85.5 & 8.9 & 65.1 & 32.7 & 11.6 \\
J020215$+$131749 & 14.3 & 0.7 & 140.1 & 14.3 & 40.1 & 30.3 & 11.4 \\
J021601$-$010312 & 3.1 & 0.2 & 43.8 & 5.0 & 27.2 & 22.8 & 11.3 \\
J022229$+$002900 & 2.2 & 0.2 & 6.9 & 4.9 & 27.4 & 23.2 & 11.0 \\
J024750$+$004718 & 6.9 & 0.4 & 167.9 & 17.1 & 185.6 & 28.1 & 11.7 \\
J025220$-$004343 & 2.1 & 0.2 & 47.7 & 5.6 & 92.3 & 14.3 & 11.5 \\
J025958$-$003622 & 25.9 & 1.3 & 327.7 & 33.0 & 267.7 & 40.6 & 11.7 \\
J031036$+$000817 & 5.8 & 0.3 & 77.9 & 8.3 & 169.7 & 26.0 & 11.5 \\
J031345$-$010517 & 4.1 & 0.3 & 54.8 & 6.2 & 106.9 & 16.4 & 11.4 \\
J032641$+$004847 & 2.7 & 0.2 & 27.7 & 3.8 & 54.9 & 41.4 & 11.3 \\
J033206$+$011048 & 3.6 & 0.2 & 50.2 & 12.8 & 119.1 & 30.7 & 11.5 \\
J033918$-$011424 & 2.6 & 0.2 & 54.1 & 5.9 & 38.0 & 32.0 & 11.2 \\
J034742$+$010959 & 6.6 & 0.4 & 83.8 & 8.8 & 128.3 & 20.3 & 11.5 \\
J034830$-$064230 & 13.3 & 0.7 & 222.5 & 22.4 & 299.8 & 45.2 & 11.5 \\
J040210$-$054630 & 1.8 & 0.2 & 28.3 & 7.0 & 18.8 & 16.8 & 11.0 \\
J073219$+$380508 & 10.1 & 0.5 & 144.1 & 14.6 & 191.5 & 28.9 & 11.4 \\
J074936$+$333716 & 14.5 & 0.7 & 111.6 & 11.5 & 46.1 & 34.7 & 11.7 \\
J075536$+$250846 & 9.4 & 0.5 & 253.6 & 25.6 & 238.2 & 36.0 & 11.8 \\
J080522$+$270829 & 56.1 & 2.8 & 594.7 & 59.6 & 429.5 & 64.5 & 11.7 \\
J081841$+$463505 & 16.5 & 0.8 & 263.5 & 26.5 & 311.7 & 46.9 & 11.8 \\
J082140$+$032147 & 14.6 & 0.7 & 272.2 & 27.4 & 304.3 & 45.7 & 11.7 \\
J082355$+$244830 & 29.8 & 1.5 & 405.7 & 40.7 & 291.0 & 43.8 & 12.0 \\
J084800$+$061837 & 4.1 & 0.3 & 90.2 & 9.3 & 144.0 & 21.8 & 11.4 \\
J084827$+$331643 & 21.2 & 1.1 & 316.9 & 31.8 & 327.9 & 49.3 & 11.2 \\
J085906$+$542150 & 2.2 & 0.2 & 56.6 & 6.0 & 86.3 & 65.8 & 11.0 \\
J090244$+$343000 & 7.3 & 0.4 & 141.1 & 14.3 & 198.3 & 29.9 & 11.5 \\
J090250$+$334901 & 35.5 & 1.8 & 673.2 & 67.4 & 682.3 & 102.4 & 11.6 \\
J090442$+$453317 & 5.1 & 0.3 & 59.2 & 6.3 & 178.2 & 26.8 & 11.2 \\
J090949$+$014847 & 20.4 & 1.0 & 380.5 & 38.2 & 598.5 & 89.9 & 11.8 \\
J091426$+$102409 & 11.1 & 0.6 & 220.2 & 22.2 & 312.4 & 47.0 & 11.6 \\
J092322$+$324830 & 14.3 & 0.7 & 290.6 & 29.2 & 252.2 & 37.9 & 11.4 \\
J092456$+$001829 & 23.9 & 1.2 & 534.7 & 53.6 & 678.6 & 101.9 & 11.8 \\
J092710$+$010232 & 27.0 & 1.4 & 634.7 & 63.6 & 426.7 & 64.2 & 11.8 \\
J092905$+$494059 & 12.4 & 0.6 & 222.1 & 22.3 & 318.4 & 47.8 & 11.6 \\
J093613$+$620905 & 8.0 & 0.4 & 176.7 & 17.8 & 278.5 & 41.9 & 11.7 \\
J093714$+$120019 & 19.2 & 1.0 & 351.9 & 35.4 & 415.9 & 62.5 & 11.5 \\
J094849$-$005314 & 7.8 & 0.4 & 152.3 & 15.5 & 288.3 & 43.5 & 11.7 \\
J095618$+$430727 & 3.5 & 0.2 & 36.1 & 4.0 & 15.5 & 17.3 & 11.2 \\
J100950$+$552336 & 9.1 & 0.5 & 171.5 & 17.3 & 317.9 & 47.7 & 11.6 \\
J101508$+$365818 & 6.0 & 0.3 & 138.9 & 14.2 & 212.2 & 31.9 & 11.5 \\
J101636$-$011358 & 13.7 & 0.7 & 214.7 & 21.7 & 246.8 & 37.2 & 11.5 \\
J101732$+$140436 & 4.1 & 0.3 & 70.5 & 7.6 & 100.1 & 15.4 & 11.3 \\
J102822$+$405558 & 10.3 & 0.5 & 225.4 & 22.7 & 345.5 & 51.9 & 11.7 \\
J102944$+$525143 & 9.5 & 0.5 & 150.7 & 15.2 & 179.0 & 26.9 & 11.6 \\
J104116$+$565345 & 9.2 & 0.5 & 162.7 & 42.8 & 370.5 & 110.8 & 11.6 \\
J104729$+$572842 & 6.0 & 0.3 & 89.9 & 27.1 & 282.6 & 93.3 & 11.6 \\
J104906$+$015920 & 6.6 & 0.4 & 90.9 & 9.6 & 166.7 & 25.3 & 11.5 \\
J105527$+$064015 & 5.5 & 0.3 & 105.2 & 11.0 & 173.4 & 26.2 & 11.3 \\
J110618$+$582441 & 12.1 & 0.6 & 280.9 & 28.2 & 320.5 & 48.1 & 11.3 \\
J110755$+$452809 & 20.7 & 1.0 & 294.1 & 29.6 & 333.0 & 50.0 & 12.1 \\
J110908$+$534143 & 6.7 & 0.4 & 92.2 & 9.4 & 197.8 & 29.7 & 11.4 \\
J111929$+$011117 & 10.3 & 0.5 & 206.8 & 21.0 & 245.2 & 37.0 & 11.5 \\
J112152$+$414757 & 17.8 & 0.9 & 273.9 & 27.5 & 286.2 & 43.0 & 11.7 \\
J112436$+$054053 & 11.1 & 0.6 & 206.0 & 21.0 & 342.0 & 51.5 & 11.8 \\
J112851$+$413455 & 5.4 & 0.3 & 115.8 & 11.8 & 173.1 & 26.1 & 11.3 \\
J113513$+$470821 & 22.6 & 1.1 & 346.5 & 34.8 & 331.3 & 49.8 & 11.4 \\
J113703$+$504420 & 8.8 & 0.5 & 171.4 & 17.3 & 249.7 & 37.5 & 11.4 \\
J115111$+$104710 & 13.7 & 0.7 & 165.7 & 16.9 & 260.8 & 39.3 & 11.1 \\
J115630$+$500822 & 5.9 & 0.3 & 60.3 & 6.3 & 85.7 & 13.0 & 11.3 \\
J115744$+$120750 & 11.3 & 0.6 & 247.4 & 24.9 & 319.0 & 47.9 & 11.6 \\
J120031$+$083114 & 29.9 & 1.5 & 290.6 & 29.2 & 182.6 & 27.5 & 12.0 \\
J120204$+$495112 & 10.9 & 0.6 & 138.6 & 14.0 & 105.7 & 16.0 & 11.8 \\
J120805$+$542258 & 3.1 & 0.2 & 18.8 & 2.5 & 81.7 & 12.4 & 11.3 \\
J121005$+$002640 & 23.4 & 1.2 & 477.5 & 47.9 & 639.2 & 96.0 & 11.6 \\
J122016$+$534028 & 15.3 & 0.8 & 284.6 & 28.6 & 301.4 & 45.3 & 11.7 \\
J122320$+$115931 & 15.4 & 0.8 & 241.4 & 24.3 & 307.9 & 46.3 & 11.5 \\
J122641$-$000620 & 5.6 & 0.3 & 76.6 & 8.2 & 193.0 & 29.1 & 11.7 \\
J123117$+$015430 & 4.4 & 0.3 & 28.9 & 3.7 & 72.9 & 11.2 & 11.3 \\
J123552$+$592400 & 11.7 & 0.6 & 218.0 & 21.9 & 270.7 & 40.7 & 11.5 \\
J123645$+$535901 & 7.9 & 0.4 & 143.9 & 14.5 & 203.0 & 30.5 & 11.5 \\
J124137$+$444453 & 7.0 & 0.4 & 100.8 & 10.3 & 123.4 & 18.6 & 11.3 \\
J124907$+$582729 & 3.6 & 0.2 & 49.9 & 5.2 & 24.9 & 21.1 & 11.3 \\
J125045$+$490640 & 19.1 & 1.0 & 315.5 & 31.6 & 394.7 & 59.3 & 11.9 \\
J125410$+$035951 & 20.6 & 1.0 & 356.2 & 35.8 & 547.4 & 82.2 & 11.7 \\
J125548$+$505716 & 33.7 & 1.7 & 485.4 & 48.6 & 420.4 & 63.1 & 11.7 \\
J130553$+$110319 & 4.8 & 0.3 & 134.3 & 13.7 & 94.1 & 14.3 & 11.5 \\
J130704$+$485845 & 30.4 & 1.5 & 526.8 & 52.8 & 443.1 & 66.5 & 11.5 \\
J130847$+$504259 & 15.0 & 0.8 & 308.4 & 31.0 & 406.1 & 61.0 & 11.4 \\
J130919$+$055049 & 6.6 & 0.4 & 82.8 & 8.7 & 175.1 & 26.4 & 11.7 \\
J131101$-$004215 & 10.2 & 0.5 & 87.4 & 9.3 & 43.8 & 34.4 & 11.5 \\
J131447$+$012759 & 8.4 & 0.4 & 153.2 & 15.6 & 232.1 & 35.0 & 11.9 \\
J131810$+$041929 & 119.1 & 6.0 & 1162.9 & 116.4 & 1190.4 & 178.7 & 11.9 \\
J132034$+$443649 & 12.7 & 0.6 & 238.3 & 23.9 & 321.3 & 48.3 & 11.5 \\
J133114$+$583342 & 13.4 & 0.7 & 216.7 & 21.8 & 338.3 & 50.8 & 11.7 \\
J134619$+$115204 & 11.2 & 0.6 & 150.8 & 15.4 & 180.6 & 27.2 & 11.5 \\
J134911$+$021716 & 9.1 & 0.5 & 131.9 & 13.5 & 176.9 & 26.7 & 11.6 \\
J135435$-$012213 & 17.9 & 0.9 & 384.2 & 38.6 & 459.0 & 69.0 & 11.5 \\
J135646$+$465414 & 11.5 & 0.6 & 201.2 & 20.2 & 365.4 & 54.9 & 11.8 \\
J140337$+$370355 & 27.1 & 1.4 & 557.0 & 55.8 & 369.3 & 55.5 & 12.0 \\
J141803$+$534104 & 12.4 & 0.6 & 263.9 & 26.5 & 297.4 & 44.7 & 11.5 \\
J142057$+$015232 & 7.4 & 0.4 & 100.2 & 10.4 & 138.9 & 21.0 & 11.6 \\
J142221$+$452011 & 57.1 & 2.9 & 725.5 & 72.7 & 579.8 & 87.0 & 12.0 \\
J143047$+$032330 & 6.1 & 0.3 & 100.1 & 10.4 & 165.2 & 24.9 & 11.2 \\
J143727$+$394530 & 5.6 & 0.3 & 99.2 & 10.1 & 126.7 & 19.1 & 11.2 \\
J145435$+$452856 & 4.2 & 0.2 & 17.9 & 2.5 & 38.4 & 30.7 & 11.2 \\
J150627$+$562702 & 4.5 & 0.3 & 22.7 & 5.4 & 19.0 & 15.9 & 11.2 \\
J150705$+$610919 & 13.0 & 0.7 & 266.2 & 26.7 & 384.1 & 57.7 & 11.7 \\
J151226$+$462903 & 5.0 & 0.3 & 118.6 & 12.0 & 208.2 & 31.3 & 11.5 \\
J151320$-$002551 & 7.0 & 0.4 & 37.9 & 4.5 & 5.1 & 15.4 & 11.0 \\
J151405$+$432528 & 9.5 & 0.5 & 153.2 & 15.5 & 324.9 & 48.8 & 11.7 \\
J152044$+$321440 & 20.7 & 1.0 & 457.7 & 45.9 & 486.1 & 73.0 & 11.5 \\
J152552$+$041732 & 3.8 & 0.2 & 45.0 & 5.1 & 72.9 & 11.4 & 11.1 \\
J153428$+$315314 & 32.9 & 1.7 & 566.8 & 56.8 & 557.3 & 83.7 & 11.5 \\
J154049$+$390350 & 23.8 & 1.2 & 214.9 & 21.6 & 85.1 & 12.9 & 11.8 \\
J154120$+$453619 & 26.2 & 1.3 & 118.5 & 12.0 & 76.0 & 39.4 & 11.6 \\
J154652$+$030402 & 19.6 & 1.0 & 266.0 & 26.8 & 295.9 & 44.6 & 11.6 \\
J155707$+$050530 & 10.4 & 0.5 & 174.4 & 17.6 & 193.6 & 29.3 & 11.2 \\
J155934$+$404144 & 3.4 & 0.2 & 48.9 & 5.3 & 90.1 & 13.6 & 11.5 \\
J160531$+$401741 & 9.7 & 0.5 & 190.2 & 19.2 & 199.6 & 30.0 & 11.7 \\
J161210$-$005756 & 10.8 & 0.6 & 219.4 & 22.2 & 265.8 & 40.2 & 11.7 \\
J161401$+$423721 & 12.6 & 0.6 & 241.8 & 24.3 & 368.0 & 55.3 & 11.4 \\
J163216$+$352449 & 10.2 & 0.5 & 186.0 & 18.7 & 211.0 & 31.7 & 11.8 \\
J204719$-$004931 & 12.9 & 0.7 & 208.4 & 21.0 & 183.2 & 27.8 & 11.4 \\
J205013$-$011521 & 3.3 & 0.2 & 60.9 & 6.6 & 79.1 & 12.4 & 11.3 \\
J205308$+$010937 & 5.0 & 0.3 & 56.7 & 6.2 & 167.4 & 25.5 & 11.2 \\
J210256$+$000955 & 10.1 & 0.5 & 199.1 & 20.1 & 258.2 & 39.0 & 11.6 \\
J210420$-$061840 & 4.0 & 0.3 & 61.4 & 6.7 & 140.3 & 21.5 & 11.5 \\
J211729$-$000410 & 10.6 & 0.6 & 208.4 & 21.1 & 218.4 & 33.0 & 11.6 \\
J213822$+$105132 & 5.5 & 0.3 & 62.4 & 6.6 & 237.1 & 35.8 & 11.5 \\
J213951$-$082538 & 12.8 & 0.7 & 206.2 & 20.9 & 334.0 & 50.2 & 11.5 \\
J221950$+$000125 & 8.6 & 0.5 & 91.7 & 9.6 & 74.5 & 11.9 & 11.4 \\
J222100$-$002537 & 10.0 & 0.5 & 156.4 & 15.9 & 167.7 & 25.6 & 11.5 \\
J223528$+$135812 & 12.8 & 0.7 & 271.6 & 27.3 & 386.3 & 58.1 & 11.7 \\
J233417$+$010353 & 5.2 & 0.3 & 74.8 & 8.0 & 106.6 & 16.3 & 11.6 \\
J234143$-$094048 & 9.1 & 0.5 & 137.8 & 14.1 & 193.2 & 29.1 & 11.8 \\
J235237$-$102943 & 7.2 & 0.4 & 79.2 & 30.8 & 281.5 & 153.4 & 11.7 \\
\enddata
\tablenotetext{*}{Columns: (1) Object name. (2) Flux 24 $\mu m$.  All flux units are $mJy$. (3) Flux uncertainty 24 $\mu m$.  (4) Flux 70 $\mu m$.  (5) Flux uncertainty 70 $\mu m$.  (6) Flux 160 $\mu m$.  (7) Flux uncertainty 160 $\mu m$.  (8) $log$ $L_{TIR}$ in units of $L_{\odot}$.}
\end{deluxetable}

\end{document}